\documentclass[10pt]{article}
\usepackage{multicol}
\usepackage{graphicx}
\usepackage{amsmath}
\usepackage[a4paper]{geometry}
\usepackage{hyperref}

\begin{document}

\begin{center}
{\Large \bf Centrality dependence of kinetic freeze-out
temperature and transverse flow velocity in high energy nuclear
collisions}

\vskip1.0cm

Muhammad Waqas$^{1,2,3,}${\footnote{E-mail:
waqas\_phy313@yahoo.com}}, Fu-Hu
Liu$^{1,2,}${\footnote{Corresponding author. E-mail:
fuhuliu@163.com; fuhuliu@sxu.edu.cn}}

{\small\it $^1$Institute of Theoretical Physics \& State Key
Laboratory of Quantum Optics and Quantum Optics Devices,\\ Shanxi
University, Taiyuan 030006, Shanxi, People's Republic of China

$^2$Collaborative Innovation Center of Extreme Optics, Shanxi
University,\\ Taiyuan 030006, Shanxi, People's Republic of China

$^3$School of Nuclear Science and Technology, University of
Chinese Academy of Sciences,\\ Beijing 100049, People's Republic
of China}

\end{center}

\vskip1.0cm

{\bf Abstract:} Centrality-dependent double-differential
transverse momentum spectra of charged pions, kaons, and
(anti)protons produced in mid-pseudorapidity interval in
$\sqrt{s_{NN}}=200$ GeV gold-gold and deuteron-gold collisions
with different centralities are analyzed by the blast-wave model
with Boltzmann-Gibbs statistics. Meanwhile, the mentioned spectra
in mid-rapidity interval in $\sqrt{s_{NN}}=2.76$ TeV lead-lead and
$\sqrt{s_{NN}}=5.02$ TeV proton-lead collisions with different
centralities are analyzed by the same model. The model results are
approximately in agreement with the experimental data in special
transverse momentum ranges. It is shown that with the increase of
event centrality and energy, the kinetic freeze-out temperature of
the emission source and the transverse flow velocity of the
produced particles slightly increase in some cases but they do not
give an obvious change in other cases. Meanwhile, the kinetic
freeze-out temperature (transverse flow velocity) increases
(decreases) with the increase of particle mass. The average
transverse momentum and initial temperature increase with the
increase of event centrality, collision energy, and particle mass.
This work also confirms the maximum size dependent effect, which
states that the main parameters such as the kinetic freeze-out
temperature and transverse flow velocity are mainly determined by
the heaviest nucleus from proton-nucleus to nucleus-nucleus
collisions.
\\



\vskip1.0cm
\begin{multicols}{2}

{\section{Introduction}}

The kinetic freeze-out temperature ($T_0$ or $T_{kin}$) of
emission source and the transverse flow velocity ($\beta_T$) of
produced particles are two important quantities at the stage of
kinetic freeze-out which is the last stage in high energy
proton-proton, proton-nucleus, and nucleus-nucleus
collisions~\cite{1,2,3}, where $T_0$ and $\beta_T$ reflect the
thermal motion of produced particles and the collective expansion
of emission source respectively. From the initial stage to the
last stage, the interacting system undergoes different stages
among which there is the stage of chemical freeze-out. It is
expected that the freeze-out parameters are event centrality and
collision energy dependent due to the fact that the violent degree
of impact is related to the amount of energy deposition which
results in given excitation and expansion degrees of the system.
In particular, the event centrality dependent freeze-out
parameters at given collision energy can be studied in
proton-nucleus and nucleus-nucleus collisions. Although, the
proton-proton collisions are not considered in the present work,
the centrality of proton-proton collisions can be also determined
by particle multiplicity.

As the result of soft excitation process, the transverse momentum
($p_T$) spectra of charged particles in low $p_T$ region contain
information of $T_0$ and $\beta_T$~\cite{1,2,3}. There are
multiple methods to extract $T_0$ and $\beta_T$. These methods
include the blast-wave model with Boltzmann-Gibbs~\cite{1,2,3} or
Tsallis statistics~\cite{4}, the alternative method using the
Boltzmann~\cite{2,5,6,7,8,9,10,11} or Tsallis
distribution~\cite{12,13}, etc. The Boltzmann-Gibbs statistics and
Boltzmann distribution are our preferred methods due to their
similarity with the ideal gas model in thermodynamics.

We can select the methods that used the Boltzmann-Gibbs statistics
and Boltzmann distribution~\cite{1,2,3} to describe the spectra in
low $p_T$ region which is less than 2--3 GeV/$c$ in peripheral
collisions and 4.5 GeV/$c$ or a little more in central collisions.
However, these methods are not suitable for the spectra in high
$p_T$ region which needs the description of other methods such as
the Hagedorn function~\cite{14,15} due to the contribution of hard
scattering process. As a probability density function, the
Hagedorn function can contribute in both the low and high $p_T$
regions. That is, except for the disengaging of $T_0$ and
$\beta_T$ in the extraction process, we should exclude the
contribution of the hard process in low $p_T$ region.
Comparatively higher values for $T_0$ and $\beta_T$ will be
obtained, if the hard process available in the Hagedorn
model~\cite{14} is included in low $p_T$ region where it's
relative fraction is small and the departure caused by the hard
process can be neglected as well. In other words, we would like to
say that we can neglect contribution of hard process in low $p_T$
region, when extracting $T_0$ and $\beta_T$ parameters.

In the present work, the centrality-dependent double-differential
transverse momentum spectra of charged pions produced in high
energy nuclear collisions will be analyzed by the blast-wave model
with Boltzmann-Gibbs statistics~\cite{1,2,3}. The model results
are compared with the experimental data measured by the PHENIX
Collaboration in mid-pseudorapidity interval in gold-gold
(Au-Au)~\cite{16} and deuteron-gold ($d$-Au)~\cite{17} collisions
at $\sqrt{s_{NN}}=200$ GeV with different centralities at the
Relativistic Heavy Ion Collider (RHIC), and lead-lead (Pb-Pb)
collisions at $\sqrt{s_{NN}}=2.76$ TeV~\cite{18} and proton-lead
($p$-Pb) collisions at $\sqrt{s_{NN}}=5.02$ TeV~\cite{19} by the
ALICE Collaboration in mid-rapidity interval with different
centralities at the Large Hadron Collider (LHC).

The remainder of this paper is structured as follows. The method
is shortly described in Section 2. Results and discussion are
given in Section 3. In Section 4, we summarize our main
observations and conclusions
\\

{\section{The method}}

The $p_T$ spectra of charged particles produced in high energy
collisions have complex structures. To describe the $p_T$ spectra,
it is not enough to use only one probability density function,
though this function can be of various forms. In particular, the
maximum $p_T$ reaches 100 GeV/$c$ in collisions at the
LHC~\cite{20}. The model analysis has observed several $p_T$
regions~\cite{21} which include the first region with $p_T<4$--6
GeV/$c$, the second region with 4--6 GeV/$c<p_T<17$--20 GeV$/c$,
and the third region with $p_T>17$--20 GeV/$c$. At the RHIC, the
boundaries of different $p_T$ regions are slightly lower. It is
expected that different $p_T$ regions correspond to different
interacting mechanisms. Even for the same $p_T$ region, different
explanations are existed due to different model methods and
microcosmic pictures.

According to ref.~\cite{21}, different $p_T$ regions reflect
different whole features of fragmentation and hadronization of
partons through the string dynamics. In the first $p_T$ region,
the effects and changes by the medium take part in the main role.
However, it appears weakly in the second $p_T$ region. Meanwhile,
the nuclear transparency results in negligible influence of the
medium in the third $p_T$ region. From the number of strings point
of view, the second $p_T$ region is expected to have the maximum
number of strings, which results in fusion and creation of strings
and collective behavior of partons. Through string fusion, the
second $p_T$ region is proposed as a possible area of Quark-Gluon
Plasma (QGP). While, the first $p_T$ region has the minimum number
of strings and maximum number of hadrons due to direct
hadronization of the low energy strings into mesons~\cite{21}.

We have used the idea of multiple $p_T$ regions and our
explanation in the following paragraphs is somehow different from
that in ref.~\cite{21}. We regard the first $p_T$ region as the
contribution region of soft excitation process. The second and
third $p_T$ regions are regarded as the contribution regions of
hard and very-hard (VH) scattering processes respectively.
Considering the contribution region ($p_T<0.2$--0.3 GeV/$c$) of
very-soft (VS) excitation process due to resonant production of
charged pions in some cases, we have one more $p_T$ region. The
four $p_T$ regions can be described by different components in a
unified superposition. To structure the unified superposition, we
have two methods. The first method is the common method of
overlapping of the contribution regions of different components,
while the second method is the Hagedorn model~\cite{14} which
doesn't include this overlapping.

Let $f_S(p_T)$, $f_H(p_T)$, $f_{VS}(p_T)$, and $f_{VH}(p_T)$
denote the probability density functions contributed by the soft,
hard, very-soft, and very-hard components, respectively, where
$f_{VS}(p_T)$ and $f_{VH}(p_T)$ are assumed to have the same forms
as $f_S(p_T)$ and $f_H(p_T)$ with smaller and larger parameters
respectively. Then, according to the first method, we can
structure the unified superposition to be
\begin{align}
f_0(p_T)=& k_{VS}f_{VS}(p_T) +kf_S(p_T) \nonumber\\
& +(1-k-k_{VS}-k_{VH})f_H(p_T) \nonumber\\
& +k_{VH}f_{VH}(p_T),
\end{align}
where $k_{VS}$, $k_{VH}$, and $k$ denote the contribution
fractions of the very-soft, very-hard, and soft components
respectively.

According to the Hagedorn's model~\cite{14}, we can use the usual
step function to structure the unified superposition. That is
\begin{align}
f_0(p_T)=& A_{VS}\theta(p_{VS}-p_T) f_{VS}(p_T) \nonumber\\
& +A_S\theta(p_T-p_{VS})\theta(p_1-p_T)f_S(p_T) \nonumber\\
& +A_H\theta(p_T-p_1)\theta(p_{VH}-p_T)f_H(p_T) \nonumber\\
& +A_{VH}\theta(p_T-p_{VH})f_{VH}(p_T),
\end{align}
where $A_{VS}$, $A_S$, $A_H$, and $A_{VH}$ are constants which
result in the two contiguous components to be equal to each other
at $p_T=p_{VS}$, $p_1$, and $p_{VH}$ respectively. In particular,
$p_1$ and $p_{VH}$ correspond to 4--6 GeV/$c$ and 17--20 GeV/$c$
in ref.~\cite{14} respectively, though the real values may be
different from them. Meanwhile, the real values are possibly event
centrality and collision energy dependent.

In most cases, the contributions of very-soft and very-hard
components can be neglected. Or, the two contributions can be
included in soft and hard components respectively. Thus, Eqs. (1)
and (2) are simplified to
\begin{align}
f_0(p_T)=kf_S(p_T)+(1-k)f_H(p_T)
\end{align}
and
\begin{align}
f_0(p_T)= A_S\theta(p_1-p_T)f_S(p_T)+A_H\theta(p_T-p_1)f_H(p_T)
\end{align}
respectively. The two simplified functions are the same to our
recent work~\cite{22} which studies the possible scenarios for
single, double, or multiple kinetic freeze-out in high energy
collisions, though $p_T$ spectra of different types of particles
produced in central and peripheral nuclear collisions and
proton-proton collisions are analyzed. Various potential functions
can be chosen for $f_S(p_T)$ and $f_H(f_T)$, which includes, but
are not limited to, the blast-wave model with Boltzmann-Gibbs
statistics~\cite{1,2,3} and the Hagedorn function~\cite{14,15}.

According to refs.~\cite{1,2,3}, the blast-wave model with
Boltzmann-Gibbs statistics results in the $p_T$ distribution to be
\begin{align}
f_S(p_T)= & C_0 p_T m_T \int_0^R rdr \nonumber\\
& \times I_0 \bigg[\frac{p_T \sinh(\rho)}{T_0} \bigg] K_1
\bigg[\frac{m_T \cosh(\rho)}{T_0} \bigg],
\end{align}
where $C_0$ is the normalized constant, $m_T=\sqrt{p_T^2+m_0^2}$
is the transverse mass, $m_0$ is the rest mass of the considered
particle, $r$ and $R$ are the radial position and the maximum
radial position in the thermal source, $I_0$ and $K_1$ are the
modified Bessel functions of the first and second kinds
respectively. In the modified Bessel functions, $\rho= \tanh^{-1}
[\beta(r)]$ is the boost angle, $\beta(r)= \beta_S(r/R)^{n_0}$ is
a self-similar flow profile, $\beta_S$ is the flow velocity on the
surface, $n_0=2$ as used in ref.~\cite{1}. In particular,
$\beta_T=(2/R^2)\int_0^R r\beta(r)dr =
2\beta_S/(n_0+2)=0.5\beta_S$.

The Hagedorn function~\cite{14,15} is known as an inverse
power-law~\cite{23,24,25},
\begin{align}
f_H(p_T)= Ap_T \bigg( 1+\frac{p_T}{ p_0} \bigg)^{-n},
\end{align}
where $p_0$ and $n$ are free parameters and $A$ is the
normalization constant. In
literature~\cite{26},~\cite{27,28,29,30,31}, and~\cite{32}, the
Hagedorn function are revised to
\begin{align}
f_H(p_T)=A \frac{p^2_T}{m_T} \bigg( 1+\frac{p_T}{ p_0}
\bigg)^{-n},
\end{align}
\begin{align}
f_H(p_T)=Ap_T \bigg( 1+\frac{p^2_T}{ p^2_0} \bigg)^{-n},
\end{align}
and
\begin{align}
f_H(p_T)=A \bigg( 1+\frac{p^2_T}{ p^2_0} \bigg)^{-n},
\end{align}
respectively, where all the three $A$, $p_0$, and $n$ are
severally different from each other.

The first method can be changed into the second method (which
results in the Hagedorn model~\cite{14}, if the contribution of
hard component in the former method can be neglected in low $p_T$
region due to its small value. If we analyze the spectra in low
$p_T$ region, the second component in Eqs. (3) and (4) should be
given up due to less contribution of $T_0$ and $\beta_T$ in hard
component. That is, we can use directly $f_S(p_T)$ from Eq. (5)
which also includes the contribution of very-soft component that
comes from resonance decays if available in the data. In this
work, the contribution of hard component in low $p_T$ region if
available is not excluded in the extraction of $T_0$ and
$\beta_T$. This treatment causes a slight increase in $T_0$ and/or
$\beta_T$ in which the relative increase is neglected due to small
value ($<5$\%)~\cite{33}.

Although we use only Eq. (5) but Eqs. (3) and (4) are kept to show
a method for further analysis if necessary. The only use Eq. (5)
in this paper means that the fraction of hard component is zero in
low $p_T$ region. In fact, it is right to exclude the contribution
of hard component in low $p_T$ region. As probability density
functions, the integrals of Eqs. (3) and (4) [$f_0(p_T)$] are
normalized to 1 respectively. Meanwhile, each component
[$f_S(p_T)$ or $f_H(p_T)$] in Eqs. (3) and (4) is also normalized
to 1 due to it being probability density function. When we compare
$f_0(p_T)$ with experimental data, we have three main relations,
$(1/2\pi p_T)d^2N/dp_Tdy=(1/2\pi p_T)N_0f_0(p_T)/dy$,
$d^2N/dp_Tdy=N_0f_0(p_T)/dy$, and $dN/dp_T=N_0f_0(p_T)$ according
to different forms of cited data, where $N$ and $N_0$ denote the
particle number and normalization constant respectively. In some
cases, $N$ ($N_0$) can be replaced by the cross-section $\sigma$
(normalization constant $\sigma_0$) if necessary.

From the above description, one can see that the method used in
this paper is quite well known. Such kind of works of fitting the
$p_T$ spectra to thermal model or Hagedorn function have been done
by many for decades. In particular, the low $p_T$ hadrons produced
in nuclear collisions are described by hydrodynamical models quite
successfully~\cite{33a,33b,33c}. However, we would like to point
out that what we will report in the following section is a more
extensive application of the thermal model in high energy nuclear
collisions at the RHIC and LHC. Meanwhile, the thermal model is
more simpler in extracting the thermal parameters, while
hydrodynamical models describe advantageously the evolution
process of collision system~\cite{33a,33b,33c}. Based on the
thermal model, the centrality dependences of the kinetic
freeze-out temperature $T_0$, transverse flow velocity $\beta_T$,
average transverse momentum $\langle p_T\rangle$, and initial
temperature $T_i$ for the emissions of identified particles, as
well as the respective weighted averages are then obtained from
systemizing the available experimental data. This systematic
analysis on the centrality dependences of multiple parameters is a
new attempt for us.
\\

{\section{Results and discussion}}

{\subsection{Comparison with experimental data}}

Figure 1 presents the event centrality dependent
double-differential $p_T$ spectra, $(1/2\pi p_T)d^2N/dp_Tdy$, of
(a)(b) $\pi^-$, (c)(d) $K^-$, and (e)(f) $\bar p$ produced in
mid-pseudorapidity interval ($|\eta|<0.35$) in (a)(c)(e) Au-Au and
(b)(d)(f) $d$-Au collisions at $\sqrt{s_{NN}}=200$ GeV at the
RHIC, where $y$ denotes rapidity and the mid-pseudorapidity
interval is decided by the PHENIX experiment~\cite{16,17} which we
cited. The symbols represent the experimental data measured by the
PHENIX Collaboration~\cite{16,17}. The spectra in different
centrality classes are scaled (multiplied) by different amounts
marked in the panels, where the centrality classifications for
Au-Au and $d$-Au collisions are different. The solid curves are
our fitted results by using separately the blast-wave model with
Boltzmann-Gibbs statistics, Eq. (5), where all the data points in
the figure are used for fitting, though the low $p_T$ range is
satisfied primarily. The dashed curves are our simultaneous fit by
the model, which will be discussed later. As a result, a special
$p_T$ range can be obtained, beyond which Eq. (5) does not work
and $f_H(p_T)$ in Eq. (3) or (4) is needed. Corresponding to
panels (a)--(f), the results of data/fit for the separate
(simultaneous) fit are presented in panels (a$'$)--(f$'$)
[(a$''$)--(f$''$)] respectively to monitor the departure of the
separate (simultaneous) fit from data. In each fitting, the method
of least squares is used in a special $p_T$ range to obtain the
best values of parameters. The values of free parameters ($T_0$
and $\beta_T$), normalization constant ($N_0$), $\chi^2$, and
number of degree of freedom (ndof) are listed in Table 1, where
the substantial event centralities and derived parameters which
will be discussed later are listed together. In particular, $N_0$
satisfies $(1/2\pi p_T)d^2N/dp_Tdy=(1/2\pi p_T)N_0f_S(p_T)/dy$.
One can see that the model results describe approximately the
PHENIX data in special $p_T$ ranges in high energy nuclear
collisions at the RHIC.

In particular, the special $p_T$ range is $0\sim2$--3 GeV/$c$ in
peripheral collisions and $0\sim4.5$ GeV/$c$ or a little more in
central collisions. This difference is caused by the fact that the
multiple scatterings in peripheral collisions are less than those
in central collisions due to less participant region in peripheral
collisions, where the spectator in peripheral collisions has less
influences. The special $p_T$ range for strange particle is
slightly narrower than that for non-strange particle. This
difference is caused by the slightly less collision cross-section
and then less frequency of multiple scatterings for strange
particle $K^-$ than for non-strange particles $\pi^-$ and $\bar
p$, which results in the special $p_T$ range for strange particle
to be slightly narrower. In addition, although Au-Au collisions
have the same $\sqrt{s_{NN}}$ as $d$-Au collisions in Fig. 1,
different sizes of participant regions for the two systems also
affect the slopes of curves due to different frequencies of
multiple scatterings. Usually, larger participant region and then
more multiple scatterings result in more gentle curve. However,
this influence is small due to the maximum size dependent
effect~\cite{33}.

\begin{figure*}[htbp]
\vskip-0cm
\begin{center}
\includegraphics[width=15.cm]{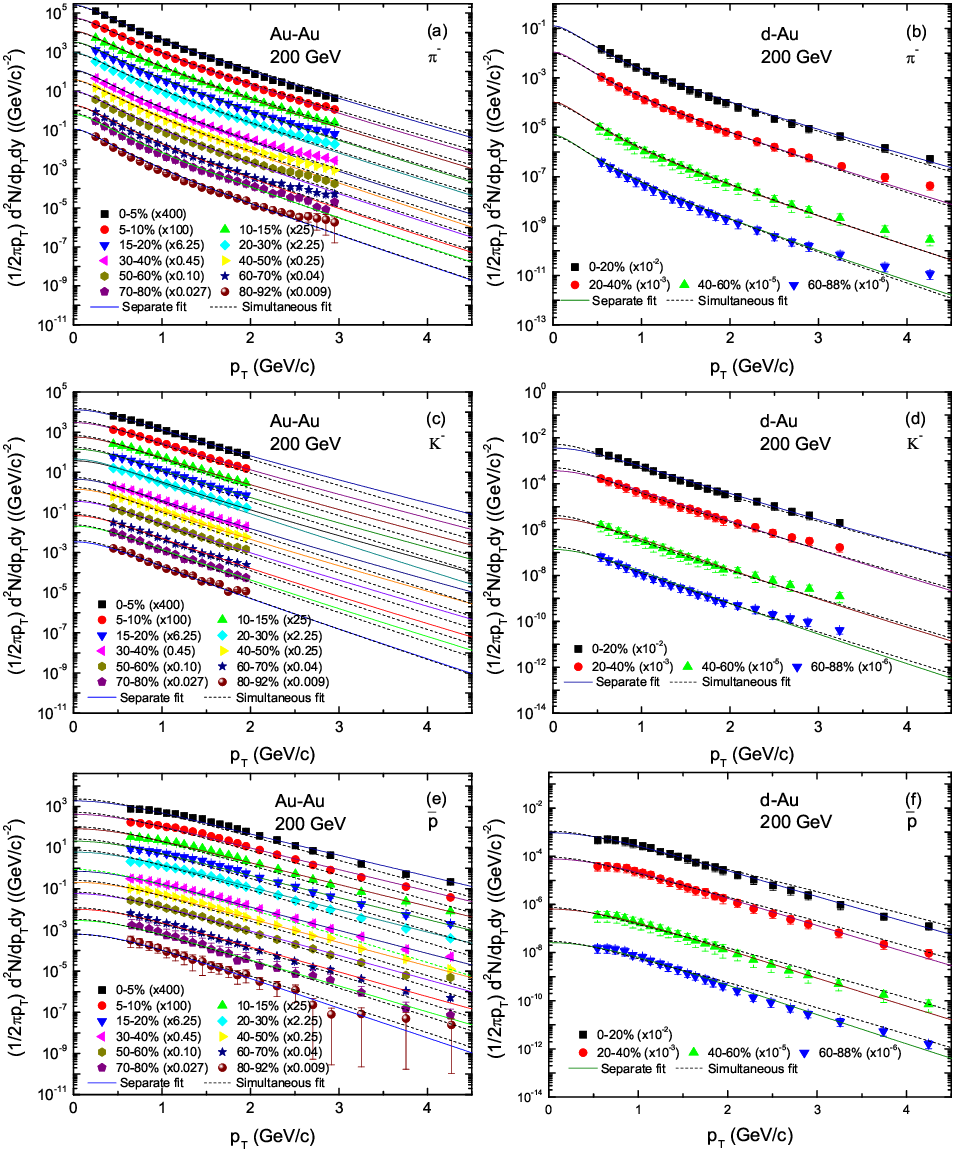}
\end{center}
Fig. 1. Centrality dependent $(1/2\pi p_T)d^2N/dp_Tdy$ of (a)(b)
$\pi^-$, (c)(d) $K^-$, and (e)(f) $\bar p$ produced in
$|\eta|<0.35$ in (a)(c)(e) Au-Au and (b)(d)(f) $d$-Au collisions
at $\sqrt{s_{NN}}=200$ GeV. The symbols represent the experimental
data measured by the PHENIX Collaboration~\cite{16,17}. The solid
curves are our fitted results by using separately the blast-wave
model with Boltzmann-Gibbs statistics, Eq. (5), while the dashed
curves are those by using simultaneously the model. The spectra in
different centrality classes are scaled by different amounts
marked in the panels and all the data points in the figure are
used for fitting.
\end{figure*}

\begin{figure*}[!htb]
\vskip-0cm
\begin{center}
\includegraphics[width=15.cm]{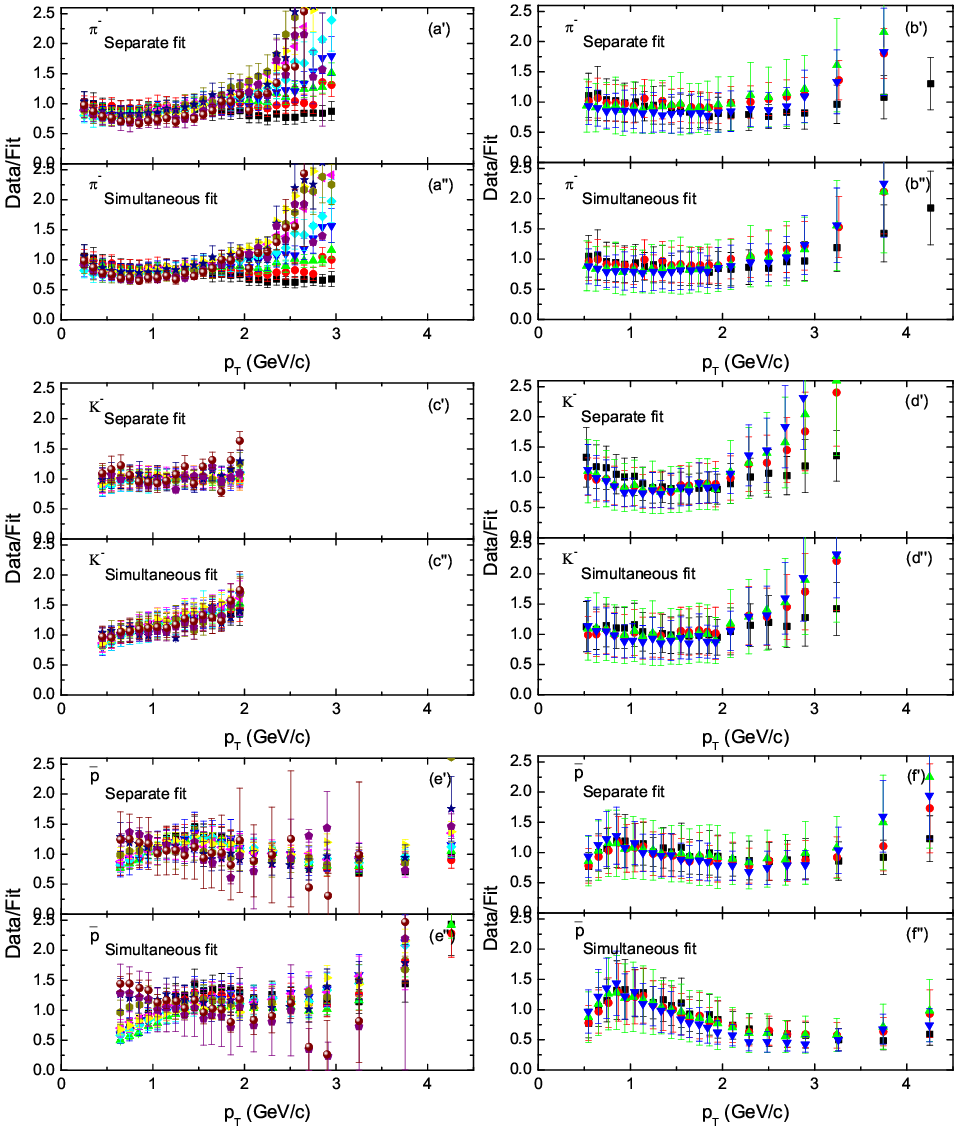}
\end{center}
Fig. 1. Continued. Corresponding to panels (a)--(f), the results
of data/fit for the separate (simultaneous) fit are presented in
panels (a$'$)--(f$'$) [(a$''$)--(f$''$)] respectively to monitor
the departure of the separate (simultaneous) fit from data. The
error bars in the data/fit are calculated according to the error
transfer formula in which only the statistical errors are
considered.
\end{figure*}

\begin{table*}[!htb]
{\small Table 1. Values of $T_0$, $\beta_T$, $\langle p_T\rangle$,
$T_i$, $N_0$, $\chi^2$, and ndof corresponding to the solid curves
in Fig. 1. \vspace{-0.25cm}
\begin{center}
\scriptsize
\begin{tabular}{ccccccccccc}\\ \hline\hline
Figure & Particle & Centrality & $T_0$ (GeV) & $\beta_T$ ($c$) &
$\langle p_T\rangle$ (GeV/$c$) & $T_i$ (GeV) &
 $N_0$ & $\chi^2$ & ndof \\ \hline
Fig. 1  &$\pi^-$   & 0--5\%   & $0.140\pm0.004$ & $0.368\pm0.004$ & $0.467\pm0.023$ & $0.412\pm0.021$ & $46.00\pm2.50$  & 34  & 25\\
Au-Au   &          & 5--10\%  & $0.138\pm0.005$ & $0.365\pm0.005$ & $0.458\pm0.022$ & $0.403\pm0.021$ & $39.00\pm2.00$  & 9   & 25\\
200 GeV &          & 10-15\%  & $0.136\pm0.003$ & $0.364\pm0.005$ & $0.451\pm0.022$ & $0.397\pm0.020$ & $32.00\pm1.40$  & 57  & 25\\
        &          & 15--20\% & $0.132\pm0.004$ & $0.366\pm0.006$ & $0.440\pm0.022$ & $0.388\pm0.019$ & $33.60\pm2.00$  & 39  & 25\\
        &          & 20--30\% & $0.130\pm0.004$ & $0.364\pm0.004$ & $0.433\pm0.022$ & $0.381\pm0.019$ & $25.30\pm2.00$  & 79  & 25\\
        &          & 30--40\% & $0.128\pm0.004$ & $0.365\pm0.006$ & $0.428\pm0.021$ & $0.376\pm0.019$ & $18.00\pm0.60$  & 111 & 25\\
        &          & 40--50\% & $0.127\pm0.003$ & $0.360\pm0.005$ & $0.421\pm0.022$ & $0.369\pm0.018$ & $8.95\pm0.44$   & 286 & 25\\
        &          & 50--60\% & $0.126\pm0.003$ & $0.365\pm0.004$ & $0.420\pm0.021$ & $0.371\pm0.019$ & $5.80\pm0.27$   & 111 & 25\\
        &          & 60--70\% & $0.125\pm0.005$ & $0.365\pm0.004$ & $0.419\pm0.020$ & $0.368\pm0.018$ & $2.70\pm0.30$   & 91  & 25\\
        &          & 70--80\% & $0.123\pm0.004$ & $0.364\pm0.006$ & $0.412\pm0.021$ & $0.362\pm0.018$ & $1.40\pm0.07$   & 92  & 25\\
        &          & 80-92\%  & $0.120\pm0.003$ & $0.367\pm0.005$ & $0.406\pm0.020$ & $0.357\pm0.018$ & $0.69\pm0.02$   & 293 & 25\\
\cline{2-10}
        &  $K^-$   & 0--5\%   & $0.180\pm0.005$ & $0.346\pm0.006$ & $0.702\pm0.035$ & $0.599\pm0.030$ & $6.85\pm0.25$   & 1  & 13\\
        &          & 5--10\%  & $0.176\pm0.006$ & $0.345\pm0.008$ & $0.688\pm0.034$ & $0.587\pm0.029$ & $6.02\pm0.27$   & 1  & 13\\
        &          & 10--15\% & $0.174\pm0.006$ & $0.344\pm0.006$ & $0.682\pm0.034$ & $0.581\pm0.029$ & $4.54\pm0.23$   & 5  & 13\\
        &          & 15--20\% & $0.172\pm0.004$ & $0.340\pm0.006$ & $0.671\pm0.034$ & $0.571\pm0.029$ & $4.48\pm0.24$   & 1  & 13\\
        &          & 20--30\% & $0.170\pm0.007$ & $0.338\pm0.007$ & $0.663\pm0.033$ & $0.564\pm0.028$ & $3.20\pm0.27$   & 2  & 13\\
        &          &30--40\%  & $0.169\pm0.004$ & $0.337\pm0.005$ & $0.659\pm0.032$ & $0.560\pm0.028$ & $1.92\pm0.10$   & 1  & 13\\
        &          &40--50\%  & $0.167\pm0.005$ & $0.335\pm0.006$ & $0.651\pm0.033$ & $0.553\pm0.028$ & $1.07\pm0.07$   & 2  & 13\\
        &          &50--60\%  & $0.164\pm0.004$ & $0.330\pm0.007$ & $0.637\pm0.032$ & $0.540\pm0.027$ & $0.64\pm0.06$   & 3  & 13\\
        &          &60--70\%  & $0.160\pm0.004$ & $0.327\pm0.005$ & $0.623\pm0.031$ & $0.528\pm0.026$ & $0.30\pm0.03$   & 4  & 13\\
        &          &70-80\%   & $0.156\pm0.004$ & $0.326\pm0.008$ & $0.611\pm0.031$ & $0.517\pm0.026$ & $0.13\pm0.02$   & 49 & 13\\
        &          &80--92\%  & $0.150\pm0.006$ & $0.317\pm0.007$ & $0.585\pm0.029$ & $0.494\pm0.025$ & $0.06\pm0.03$   & 49 & 13\\
\cline{2-10}
        & $\bar p$ &0--5\%    & $0.208\pm0.006$ & $0.333\pm0.005$ & $0.939\pm0.047$ & $0.786\pm0.039$ & $1.97\pm0.10$   & 36  & 19\\
        &          &5--10\%   & $0.206\pm0.005$ & $0.326\pm0.004$ & $0.926\pm0.046$ & $0.775\pm0.039$ & $1.71\pm0.11$   & 25  & 19\\
        &          &10-15\%   & $0.204\pm0.005$ & $0.325\pm0.007$ & $0.919\pm0.046$ & $0.769\pm0.038$ & $1.34\pm0.04$   & 63  & 19\\
        &          &15-20\%   & $0.203\pm0.004$ & $0.322\pm0.004$ & $0.912\pm0.046$ & $0.762\pm0.038$ & $0.13\pm0.06$   & 7   & 19\\
        &          &20--30\%  & $0.200\pm0.005$ & $0.319\pm0.004$ & $0.898\pm0.045$ & $0.751\pm0.038$ & $0.13\pm0.11$   & 22  & 19\\
        &          &30-40\%   & $0.198\pm0.004$ & $0.312\pm0.004$ & $0.881\pm0.044$ & $0.735\pm0.037$ & $0.60\pm0.02$   & 8   & 19\\
        &          &40--50\%  & $0.196\pm0.006$ & $0.308\pm0.003$ & $0.870\pm0.044$ & $0.725\pm0.036$ & $0.31\pm0.03$   & 22  & 19\\
        &          &50--60\%  & $0.191\pm0.003$ & $0.305\pm0.003$ & $0.852\pm0.043$ & $0.710\pm0.036$ & $0.21\pm0.02$   & 33  & 19\\
        &          &60--70\%  & $0.188\pm0.004$ & $0.304\pm0.004$ & $0.843\pm0.042$ & $0.702\pm0.035$ & $0.090\pm0.010$ & 35  & 19\\
        &          &70--80\%  & $0.184\pm0.003$ & $0.294\pm0.003$ & $0.818\pm0.042$ & $0.679\pm0.034$ & $0.036\pm0.002$ & 68  & 19\\
        &          &80--92\%  & $0.162\pm0.006$ & $0.285\pm0.005$ & $0.749\pm0.038$ & $0.620\pm0.031$ & $0.020\pm0.002$ & 194 & 19\\
\hline
Fig. 1  & $\pi^-$  & 0--20\%  & $0.120\pm0.003$ & $0.443\pm0.003$ & $0.504\pm0.033$ & $0.464\pm0.023$ & $0.86\pm0.04$   & 10 & 21\\
$d$-Au  &          & 20--40\% & $0.117\pm0.004$ & $0.436\pm0.003$ & $0.481\pm0.022$ & $0.440\pm0.022$ & $0.70\pm0.08$   & 10 & 21\\
200 GeV &          & 40--60\% & $0.112\pm0.004$ & $0.435\pm0.004$ & $0.462\pm0.023$ & $0.422\pm0.021$ & $0.62\pm0.06$   & 7  & 21\\
        &          & 60--88\% & $0.109\pm0.003$ & $0.432\pm0.004$ & $0.447\pm0.022$ & $0.406\pm0.020$ & $0.28\pm0.02$   & 15 & 21\\
\cline{2-10}
        & $K^-$    & 0--20\%  & $0.239\pm0.006$ & $0.254\pm0.005$ & $0.752\pm0.038$ & $0.635\pm0.032$ & $0.100\pm0.010$ & 8  & 18\\
        &          & 20-40\%  & $0.235\pm0.004$ & $0.250\pm0.005$ & $0.738\pm0.037$ & $0.622\pm0.031$ & $0.090\pm0.005$ & 9  & 18\\
        &          & 40-60\%  & $0.230\pm0.005$ & $0.248\pm0.003$ & $0.724\pm0.036$ & $0.610\pm0.031$ & $0.073\pm0.005$ & 6  & 18\\
        &          & 60-88\%  & $0.218\pm0.004$ & $0.247\pm0.006$ & $0.694\pm0.035$ & $0.584\pm0.029$ & $0.030\pm0.003$ & 18 & 18\\
\cline{2-10}
        &$\bar p$  & 0--20\%  & $0.262\pm0.004$ & $0.238\pm0.006$ & $0.935\pm0.047$ & $0.777\pm0.039$ & $0.042\pm0.004$ & 6  & 21\\
        &          & 20--40\% & $0.260\pm0.005$ & $0.230\pm0.005$ & $0.922\pm0.046$ & $0.765\pm0.038$ & $0.033\pm0.002$ & 6  & 21\\
        &          & 40--60\% & $0.250\pm0.006$ & $0.228\pm0.005$ & $0.896\pm0.045$ & $0.743\pm0.037$ & $0.026\pm0.003$ & 4  & 21\\
        &          & 60--88\% & $0.245\pm0.004$ & $0.215\pm0.005$ & $0.870\pm0.044$ & $0.721\pm0.036$ & $0.010\pm0.001$ & 14 & 21\\
\hline
\end{tabular}%
\end{center}}
\end{table*}

\begin{figure*}[!htb]
\vskip-0cm
\begin{center}
\includegraphics[width=15.cm]{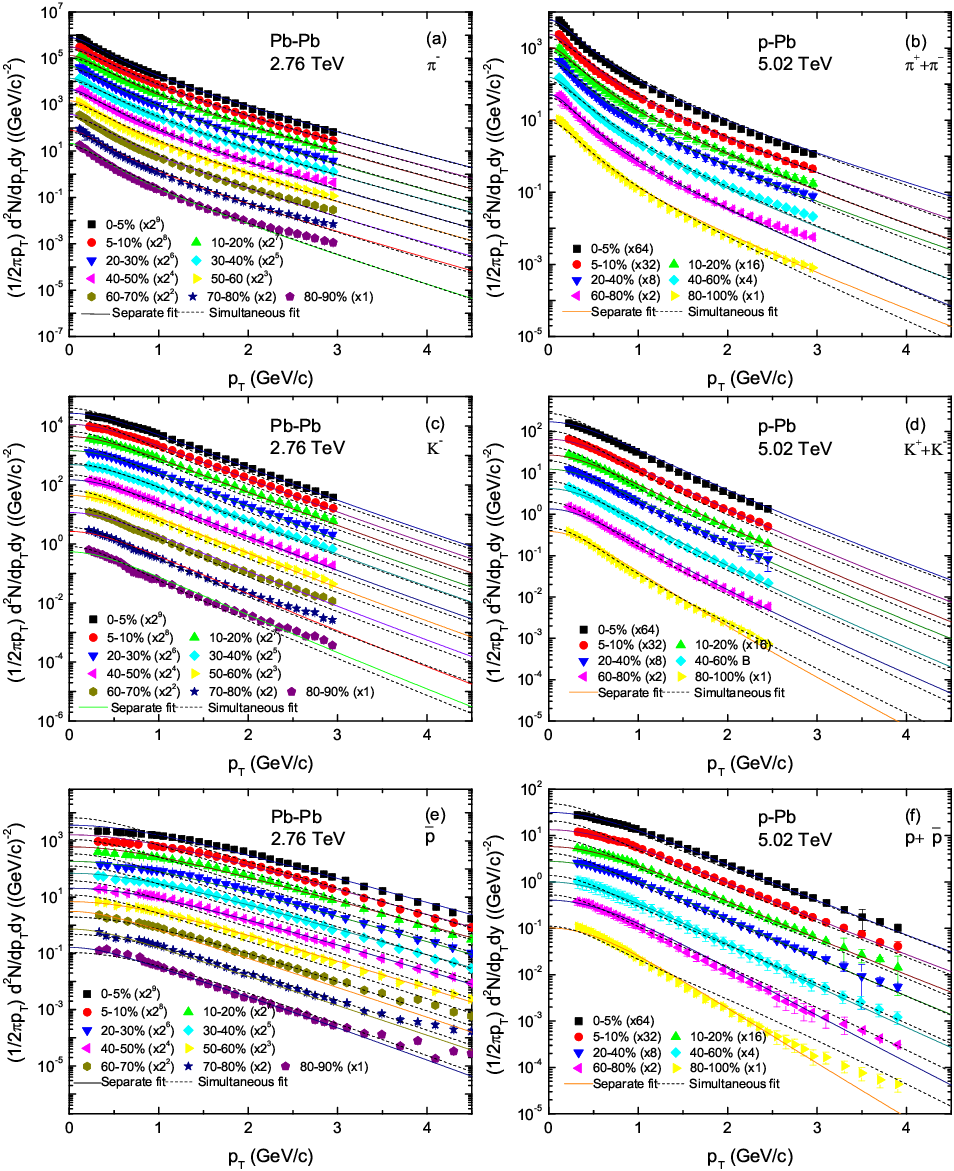}
\end{center}
Fig. 2. Same as Fig. 1, but for the spectra of (a) $\pi^-$, (c)
$K^-$, and (e) $\bar p$ produced in $|y|<0.5$ in Pb-Pb collisions
at $\sqrt{s_{NN}}=2.76$ TeV, and the spectra of (b) $\pi^++\pi^-$,
(d) $K^++K^-$, and (f) $p+\bar p$ produced in $0<y<0.5$ in $p$-Pb
collisions at $\sqrt{s_{NN}}=5.02$ TeV. The symbols represent the
experimental data measured by the ALICE
Collaboration~\cite{18,19}, where the spectra in different
centrality classes are scaled by different amounts shown in the
panels and all the data points in the figure are used for fitting.
\end{figure*}

\begin{figure*}[!htb]
\vskip-0cm
\begin{center}
\includegraphics[width=15.cm]{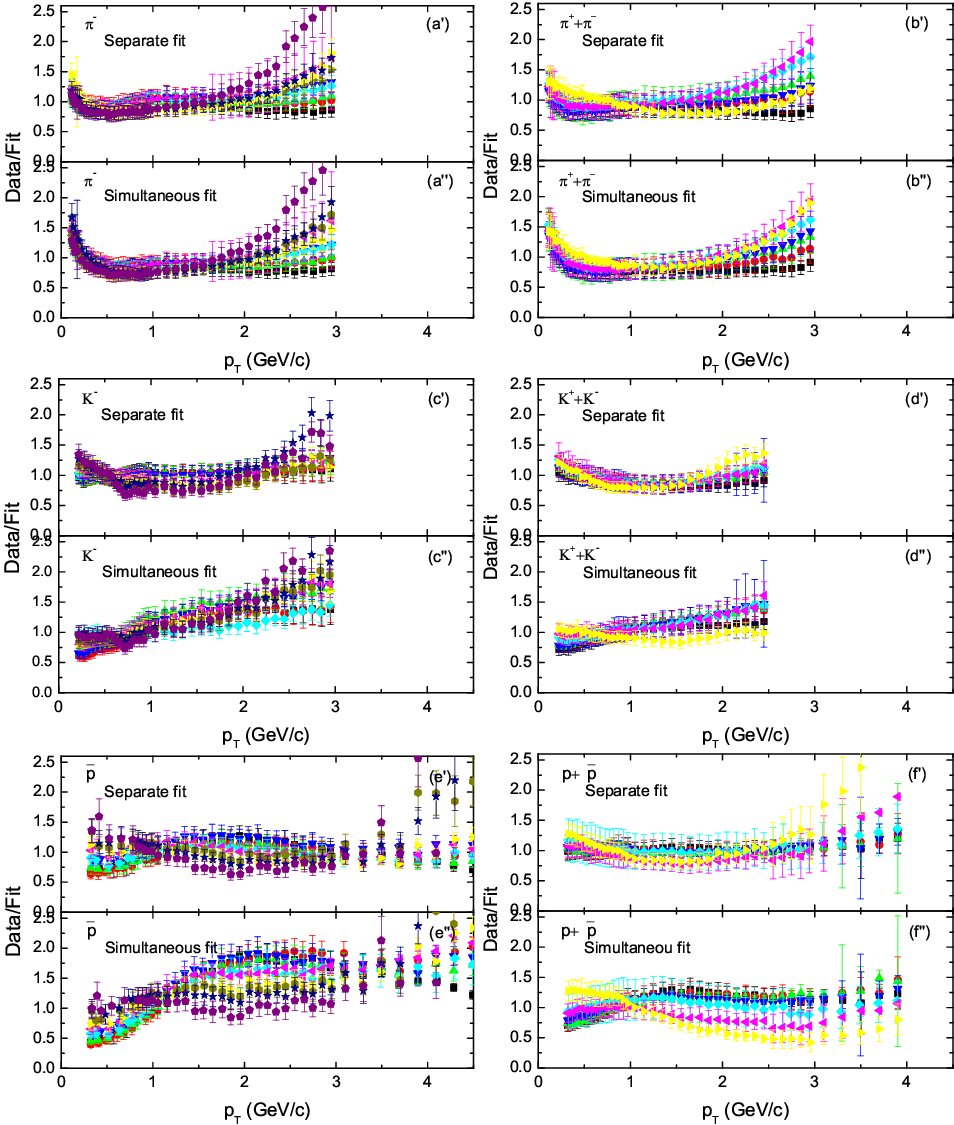}
\end{center}
Fig. 2. Continued. Corresponding to panels (a)--(f), the results
of data/fit for the separate (simultaneous) fit are presented in
panels (a$'$)--(f$'$) [(a$''$)--(f$''$)] respectively to monitor
the departure of the separate (simultaneous) fit from data. The
error bars in the data/fit are calculated according to the error
transfer formula in which only the statistical errors are
considered.
\end{figure*}

\begin{table*}[!htb]
{\small Table 2. Values of $T_0$, $\beta_T$, $\langle p_T\rangle$,
$T_i$, $N_0$, $\chi^2$, and ndof corresponding to the solid curves
in Fig. 2. \vspace{-0.25cm}
\begin{center}
\scriptsize
\begin{tabular}{cccccccccc}\\ \hline\hline
Figure & Particle & Centrality & $T_0$ (GeV) & $\beta_T$ ($c$) &
$\langle p_T\rangle$ (GeV/$c$) & $T_i$ (GeV) &
 $N_0$ & $\chi^2$ & ndof \\ \hline
Fig. 2   & $\pi^-$& 0--5\%   & $0.135\pm0.003$ & $0.430\pm0.003$ & $0.532\pm0.027$ & $0.485\pm0.024$ & $123.39\pm1.95$ & 31  & 38 \\
 Pb-Pb   &        & 5--10\%  & $0.134\pm0.004$ & $0.429\pm0.004$ & $0.526\pm0.003$ & $0.479\pm0.024$ & $105.34\pm11.72$& 8   & 38 \\
2.76 TeV &        & 10--20\% & $0.132\pm0.004$ & $0.428\pm0.005$ & $0.518\pm0.026$ & $0.471\pm0.024$ & $78.36\pm7.81$  & 15  & 38 \\
         &        & 20--30\% & $0.129\pm0.003$ & $0.427\pm0.006$ & $0.506\pm0.025$ & $0.459\pm0.023$ & $51.56\pm4.69$  & 28  & 38 \\
         &        & 30--40\% & $0.131\pm0.005$ & $0.428\pm0.005$ & $0.514\pm0.026$ & $0.468\pm0.023$ & $34.87\pm3.11$  & 20  & 38 \\
         &        & 40--50\% & $0.127\pm0.003$ & $0.426\pm0.004$ & $0.497\pm0.025$ & $0.451\pm0.023$ & $20.23\pm2.50$  & 20  & 38 \\
         &        & 50--60\% & $0.124\pm0.003$ & $0.428\pm0.004$ & $0.490\pm0.025$ & $0.445\pm0.022$ & $12.02\pm1.25$  & 67  & 38 \\
         &        & 60--70\% & $0.123\pm0.004$ & $0.426\pm0.005$ & $0.483\pm0.024$ & $0.438\pm0.022$ & $6.43\pm0.04$   & 72  & 38 \\
         &        & 70--80\% & $0.126\pm0.004$ & $0.424\pm0.005$ & $0.490\pm0.025$ & $0.444\pm0.022$ & $2.75\pm0.30$   & 83  & 38 \\
         &        & 80--90\% & $0.108\pm0.004$ & $0.428\pm0.006$ & $0.435\pm0.022$ & $0.395\pm0.020$ & $1.20\pm0.20$   & 165 & 38 \\
\cline{2-10}
         & $K^-$  & 0--5\%   & $0.289\pm0.005$ & $0.228\pm0.004$ & $0.851\pm0.043$ & $0.717\pm0.036$ & $16.80\pm0.59$  & 4   & 33 \\
         &        & 5--10\%  & $0.287\pm0.006$ & $0.227\pm0.007$ & $0.845\pm0.042$ & $0.713\pm0.036$ & $14.67\pm1.17$  & 6   & 33 \\
         &        & 10--20\% & $0.284\pm0.005$ & $0.225\pm0.006$ & $0.836\pm0.042$ & $0.705\pm0.035$ & $10.55\pm0.78$  & 11  & 33 \\
         &        & 20--30\% & $0.282\pm0.004$ & $0.227\pm0.005$ & $0.833\pm0.042$ & $0.702\pm0.035$ & $7.03\pm0.63$   & 17  & 33 \\
         &        & 30--40\% & $0.281\pm0.007$ & $0.225\pm0.005$ & $0.829\pm0.041$ & $0.699\pm0.035$ & $4.69\pm0.47$   & 14  & 33 \\
         &        & 40--50\% & $0.279\pm0.006$ & $0.222\pm0.008$ & $0.821\pm0.041$ & $0.692\pm0.035$ & $2.81\pm0.13$   & 19  & 33 \\
         &        & 50--60\% & $0.275\pm0.005$ & $0.219\pm0.008$ & $0.810\pm0.041$ & $0.682\pm0.034$ & $1.63\pm0.13$   & 41  & 33 \\
         &        & 60--70\% & $0.271\pm0.006$ & $0.217\pm0.009$ & $0.799\pm0.040$ & $0.672\pm0.034$ & $0.84\pm0.10$   & 74  & 33 \\
         &        & 70--80\% & $0.255\pm0.007$ & $0.215\pm0.010$ & $0.759\pm0.038$ & $0.638\pm0.032$ & $0.37\pm0.20$   & 105 & 33 \\
         &        & 80--90\% & $0.254\pm0.007$ & $0.209\pm0.012$ & $0.752\pm0.038$ & $0.632\pm0.032$ & $0.14\pm0.01$   & 179 & 33 \\
\cline{2-10}
         &$\bar p$& 0--5\%   & $0.443\pm0.005$ & $0.098\pm0.008$ & $1.234\pm0.062$ & $1.020\pm0.051$ & $5.27\pm0.39$   & 81  & 34 \\
         &        & 5--10\%  & $0.440\pm0.004$ & $0.050\pm0.007$ & $1.211\pm0.061$ & $1.000\pm0.050$ & $4.69\pm0.35$   & 98  & 34 \\
         &        & 10--20\% & $0.438\pm0.004$ & $0.090\pm0.007$ & $1.220\pm0.061$ & $1.008\pm0.050$ & $3.52\pm0.23$   & 61  & 34 \\
         &        & 20--30\% & $0.435\pm0.004$ & $0.060\pm0.008$ & $1.203\pm0.060$ & $0.994\pm0.050$ & $2.11\pm0.16$   & 45  & 34 \\
         &        & 30--40\% & $0.430\pm0.005$ & $0.055\pm0.013$ & $1.191\pm0.060$ & $0.984\pm0.049$ & $1.56\pm0.19$   & 26  & 34 \\
         &        & 40--50\% & $0.427\pm0.004$ & $0.026\pm0.016$ & $1.180\pm0.059$ & $0.974\pm0.049$ & $0.90\pm0.03$   & 16  & 34 \\
         &        & 50--60\% & $0.405\pm0.003$ & $0.060\pm0.012$ & $1.139\pm0.057$ & $0.940\pm0.047$ & $0.56\pm0.02$   & 48  & 34 \\
         &        & 60--70\% & $0.378\pm0.004$ & $0.046\pm0.012$ & $1.076\pm0.054$ & $0.888\pm0.044$ & $0.29\pm0.01$   & 53  & 34 \\
         &        & 70--80\% & $0.362\pm0.004$ & $0.089\pm0.004$ & $1.054\pm0.053$ & $0.870\pm0.044$ & $0.14\pm0.02$   & 72  & 34 \\
         &        & 80--90\% & $0.340\pm0.005$ & $0.080\pm0.009$ & $1.001\pm0.050$ & $0.826\pm0.041$ & $0.055\pm0.004$ & 161 & 34 \\
\hline
Fig. 2   &$\pi^++\pi^-$& 0--5\%   & $0.119\pm0.006$ & $0.469\pm0.005$ & $0.576\pm0.029$ & $0.546\pm0.027$ & $7.23\pm0.47$  & 31  & 38 \\
$p$-Pb   &             & 5--10\%  & $0.116\pm0.006$ & $0.465\pm0.005$ & $0.549\pm0.027$ & $0.518\pm0.026$ & $5.35\pm0.01$  & 11  & 38 \\
5.02 TeV &             & 10--20\% & $0.113\pm0.007$ & $0.462\pm0.006$ & $0.528\pm0.026$ & $0.495\pm0.025$ & $4.38\pm0.01$  & 14  & 38 \\
         &             & 20-40\%  & $0.112\pm0.005$ & $0.460\pm0.004$ & $0.530\pm0.027$ & $0.499\pm0.025$ & $4.06\pm0.13$  & 26  & 38 \\
         &             & 40-60\%  & $0.110\pm0.004$ & $0.457\pm0.005$ & $0.502\pm0.025$ & $0.468\pm0.023$ & $2.40\pm0.25$  & 85  & 38 \\
         &             & 60-80\%  & $0.107\pm0.005$ & $0.453\pm0.004$ & $0.480\pm0.024$ & $0.445\pm0.022$ & $1.55\pm0.06$  & 44  & 38 \\
         &             & 80-100\% & $0.100\pm0.001$ & $0.463\pm0.001$ & $0.480\pm0.024$ & $0.450\pm0.023$ & $0.60\pm0.05$  & 291 & 38 \\
\cline{2-10}
         & $K^++K^-$   & 0--5\%   & $0.293\pm0.005$ & $0.313\pm0.005$ & $0.951\pm0.048$ & $0.812\pm0.041$ & $0.98\pm0.05$  & 49 & 28 \\
         &             & 5--10\%  & $0.285\pm0.003$ & $0.310\pm0.004$ & $0.926\pm0.046$ & $0.791\pm0.040$ & $0.71\pm0.02$  & 13 & 28 \\
         &             & 10--20\% & $0.279\pm0.005$ & $0.307\pm0.006$ & $0.907\pm0.045$ & $0.774\pm0.039$ & $0.57\pm0.04$  & 57 & 28 \\
         &             & 20--40\% & $0.270\pm0.006$ & $0.312\pm0.007$ & $0.891\pm0.045$ & $0.760\pm0.038$ & $0.50\pm0.04$  & 12 & 28 \\
         &             & 40--60\% & $0.255\pm0.004$ & $0.310\pm0.005$ & $0.850\pm0.043$ & $0.725\pm0.036$ & $0.31\pm0.03$  & 70 & 28 \\
         &             & 60--80\% & $0.232\pm0.004$ & $0.329\pm0.005$ & $0.815\pm0.041$ & $0.697\pm0.035$ & $0.19\pm0.01$  & 24 & 28 \\
         &             & 80--100\%& $0.200\pm0.005$ & $0.309\pm0.006$ & $0.707\pm0.035$ & $0.600\pm0.030$ & $0.083\pm0.007$& 77 & 28 \\
\cline{2-10}
         & $p+\bar p$  & 0--5\%   & $0.355\pm0.005$ & $0.293\pm0.004$ & $1.229\pm0.061$ & $1.026\pm0.051$ & $0.34\pm0.02$  & 5  & 36 \\
         &             & 5--10\%  & $0.350\pm0.006$ & $0.290\pm0.005$ & $1.213\pm0.061$ & $1.012\pm0.051$ & $0.28\pm0.01$  & 5  & 36 \\
         &             & 10--20\% & $0.340\pm0.006$ & $0.289\pm0.006$ & $1.188\pm0.059$ & $0.992\pm0.050$ & $0.24\pm0.01$  & 14 & 36 \\
         &             & 20--40\% & $0.325\pm0.005$ & $0.282\pm0.007$ & $1.143\pm0.057$ & $0.954\pm0.048$ & $0.20\pm0.01$  & 19 & 36 \\
         &             & 40--60\% & $0.320\pm0.006$ & $0.276\pm0.005$ & $1.075\pm0.054$ & $0.897\pm0.045$ & $0.14\pm0.01$  & 35 & 36 \\
         &             & 60--80\% & $0.295\pm0.006$ & $0.227\pm0.007$ & $1.001\pm0.050$ & $0.832\pm0.042$ & $0.10\pm0.01$  & 26 & 36 \\
         &             & 80--100\%& $0.240\pm0.006$ & $0.238\pm0.006$ & $0.883\pm0.044$ & $0.732\pm0.037$ & $0.040\pm0.004$& 50 & 36 \\
\hline
\end{tabular}%
\end{center}}
\end{table*}

Figure 2 is the same as Fig. 1, but it shows the results of (a)
$\pi^-$, (c) $K^-$, and (e) $\bar p$ produced in mid-rapidity
interval $|y|<0.5$ in Pb-Pb collisions at $\sqrt{s_{NN}}=2.76$
TeV, and the results of (b) $\pi^++\pi^-$, (d) $K^++K^-$, and (f)
$p+\bar p$ produced in $0<y<0.5$ in $p$-Pb collisions at
$\sqrt{s_{NN}}=5.02$ TeV, where the positive and negative
particles in $p$-Pb collisions are not separated in experiments.
The symbols represent the experimental data measured by the ALICE
Collaboration~\cite{18,19}, where the spectra in Pb-Pb (or $p$-Pb)
collisions are scaled by different amounts for different
centrality classes shown in the panels and all the data points in
the figure are used for fitting. Corresponding to panels (a)--(f),
the results of data/fit for the separate (simultaneous) fit are
presented in panels (a$'$)--(f$'$) [(a$''$)--(f$''$)]
respectively. The related parameters are listed in Table 2, where
the existent event centralities are listed together. One can see
that the model results describe approximately the ALICE data in
special $p_T$ ranges in high energy nuclear collisions at the LHC.
The special $p_T$ range increases from $0\sim2$--3 GeV/$c$ to
$0\sim4.5$ GeV/$c$ or a little more when the event centrality
increases from periphery to center. This range for strange
particle is slightly narrower than that for non-strange particle.
The dependence of this range on energy is not obvious.

It should be noted that the uncertainties of free parameters $T_0$
and $\beta_T$ are very small due to the strict restriction for the
range of $\chi^2$. In fact, we restrict $\chi^2$ so that
$\chi^2_{\min} \leq \chi^2 \leq 1.05\chi^2_{\min}$, where
$\chi^2_{\min}$ denotes the minimum-$\chi^2$ which is obtained by
the method of least squares and which changes for each fit (each
particle in each centrality class). In the case of using weak
restrictions, for example $\chi^2_{\min} \leq \chi^2 \leq
1.10\chi^2_{\min}$, large uncertainties will be obtained, which
are not expected by us due to inaccurate determination of
parameters. In addition, Eq. (4) is not an ideal fitting function
due to fewer free parameters being used in low $p_T$ region, which
renders small variable ranges of free parameters in limited
selection. Contrarily, to give a better fit, Eq. (3) is more ideal
due to more free parameters being used in low $p_T$ region, which
renders large variable ranges of free parameters by flexible
selection. The limited selection in Eq. (4) restricts $T_0$ and
$\beta_T$ themselves.

In the fit in Figs. 1 and 2, because of only two free parameters
being used, it is possibly coincidental if $\chi^2$/ndof $<1$.
Contrarily, the case of $\chi^2$/ndof $\gg1$ is caused by the fact
that the data points in high $p_T$ region are included. Then, the
two-component functions can be used if necessary. This paper is
focused on the extraction of centrality dependence of $T_0$ and
$\beta_T$ in nuclear collisions at the top RHIC and LHC energies.
In fact, we do not need to consider the data points in high $p_T$
region. On one hand, these data points have zero contribution due
to their non-thermal processes, which should be excluded. On the
other hand, these data points have less contributions due to their
small amounts, which can be neglected. In the calculation of
$\chi^2$/ndof, we have included these data points, which causes
$\chi^2$/ndof $\gg1$.
\\

{\subsection{Trend of parameters}}

To study the dependences of $T_0$ and $\beta_T$ on the event
centrality, Figures 3 and 4 show the correlations between $T_0$
and $C$ as well as $\beta_T$ and $C$ respectively, where $C$
denotes the event centrality percentage in which 0\% centrality is
the most central collisions and 100\% centrality is the most
peripheral collisions. Different symbols represent different
parameter values listed in Tables 1 and 2. In particular, the
averages, $\langle T_0\rangle$ ($\langle \beta_T\rangle$), of
$T_0$ ($\beta_T$) weighted over yields of different particles are
shown in Fig. 3 (4) by the open circles. It can be seen that,
$T_0$ and $\beta_T$ increase slightly in some cases or almost do
not change in other cases with the increase of event centrality
from peripheral to central collisions. This difference should be
studied in the future. In particular, for $T_0$ from $K$ and $p$
spectra there is slight centrality dependence, while for $T_0$
from $\pi$ spectra at the LHC there is no obvious centrality
dependence. Except $K$ and $p$ in Au-Au collisions at 200 GeV and
$p$ in $p$-Pb collisions at 5.02 TeV, there is no dependence of
the parameter $\beta_T$ on centrality. With the increase of
particle mass, $T_0$ increases while $\beta_T$ decreases
obviously. In most cases, $T_0$ and $\beta_T$ in Pb-Pb ($p$-Pb)
collisions at the LHC are comparable with those in Au-Au ($d$-Au)
collisions at the RHIC within errors. At the same time, $T_0$ and
$\beta_T$ in Au-Au (Pb-Pb) collisions are comparable with those in
$d$-Au ($p$-Pb) collisions within errors. In short, $T_0$ and
$\beta_T$ do not decrease in general with the increase of event
centrality, collision energy, and projectile size (in the case of
the same target nucleus) within errors, and $T_0$ ($\beta_T$)
increases (decreases) obviously with the increase of particle
mass. The mass dependent $T_0$ and $\beta_T$ is a reflection of
the scenario of multiple kinetic freeze-out~\cite{22}. According
to the hydrodynamical behavior~\cite{35a}, the massive particles
are left over early due to small $\beta_T$. Meanwhile, early
emission results in high $T_0$.

We would like to point out that although we fit the three spectra
of identified particle species individually in Figs. 1 and 2 by
the solid curves, the average parameters weighted by particle
yields $N_0$ in Tables 1 and 2 should be used for the three
particles simultaneously by the dashed curves. Figures 3 and 4
show that the average parameters weighted by particle yields are
closer to those for pions due to the fact that the yield of pions
is the most at the considered energy. We notice that the average
$T_0$ over all particles in given centrality is close to or does
not exceed the chemical freeze-out temperature in general. In
addition, when we apply the weighted average parameters for kaons
and protons, we obtain relative large $\chi^2$. The dashed curves
in Figs. 1 and 2 show that both the simultaneous fits at the RHIC
and LHC are approximately and similarly successful. It should be
noted that the multiplicity weighted mean parameters are
identified with a single $\langle T_0\rangle$ and $\langle
\beta_T\rangle$ which is obtained by the fit of all particles in a
given centrality class. These mean parameters may be different
from $T_0$ and $\beta_T$ used in Refs.~\cite{2,18} due to
different restricted conditions. These restrictions include the
particle-dependent or independent $p_T$ range, unfixed or fixed
flow profile ($n_0$), and large or small $\beta_T$ change. The
present work uses the restrictions of particle-independent $p_T$
range, fixed flow profile ($n_0=2$ as used in ref.~\cite{1}), and
small $\beta_T$ change, which contains less free parameters in the
analysis and larger flow effect in peripheral collisions.

If simultaneous fit has more credibility in general, the different
values of freeze-out parameters for individual particles imply a
multiple kinetic freeze-out scenario~\cite{10,22,34,35} in terms
of detailed analysis. For particles in collisions with given
centrality, the higher the $T_0$ is, the earlier the emission of
the particles is. As one of constituents in projectile and target
participants, some protons are leading protons which are existed
in the initial state of collisions and they are emitted much
earlier than pions due to their existence before thermalization.
Even if other protons formed in the collisions are emitted
simultaneously with pions, on average, protons are emitted earlier
than pions. Generally, leading protons emitted earlier than pions
appear in the forward/backward rapidity region, while protons
emitted simultaneously with pions appear in the whole rapidity
region as pions. At the LHC, the large difference in $T_0$ for
different particles renders naturally large difference in emission
time. On the other hand, the mass-dependent multiple scenario
``shows massive particles coming out of the system earlier in time
with smaller radial flow velocities, which is hydrodynamic
behavior"~\cite{35a}. This earlier freeze-out for massive
particles appear due to the fact that they are left behind in the
system process due to low $\beta_T$ and large $m_0$, but not high
$T_0$. For collisions with different centralities, the higher the
$T_0$ is, the higher the excitation degree is. Finally, $T_0$ is a
result of competition between excitation degree and emission time.

In addition, we have used the specific profile for the transverse
flow velocity which is also used in the original blast-wave model
with Boltzmann-Gibbs statistics~\cite{1,2,3}, though the profile
is sensitive to the fit for transverse momentum spectra as
explored e.g. in refs.~\cite{18,36}. This sensitive profile
affects only the absolute sizes of $T_0$ ($\beta_T$) for emissions
of individual particles, but not the relative sizes. In other
words, this sensitive profile does not affect our conclusions on
multiple kinetic freeze-out scenario~\cite{10,22,34,35} in terms
of detailed analysis and more credible simultaneous fit in
general. It does not affect the trend on centrality dependence of
$T_0$ ($\beta_T$) too, which are discussed in Figs. 3 and 4.
Therefore, we would like not to use other specific profiles in the
present work.

\begin{figure*}[!htb]
\vskip-0cm
\begin{center}
\includegraphics[width=16.cm]{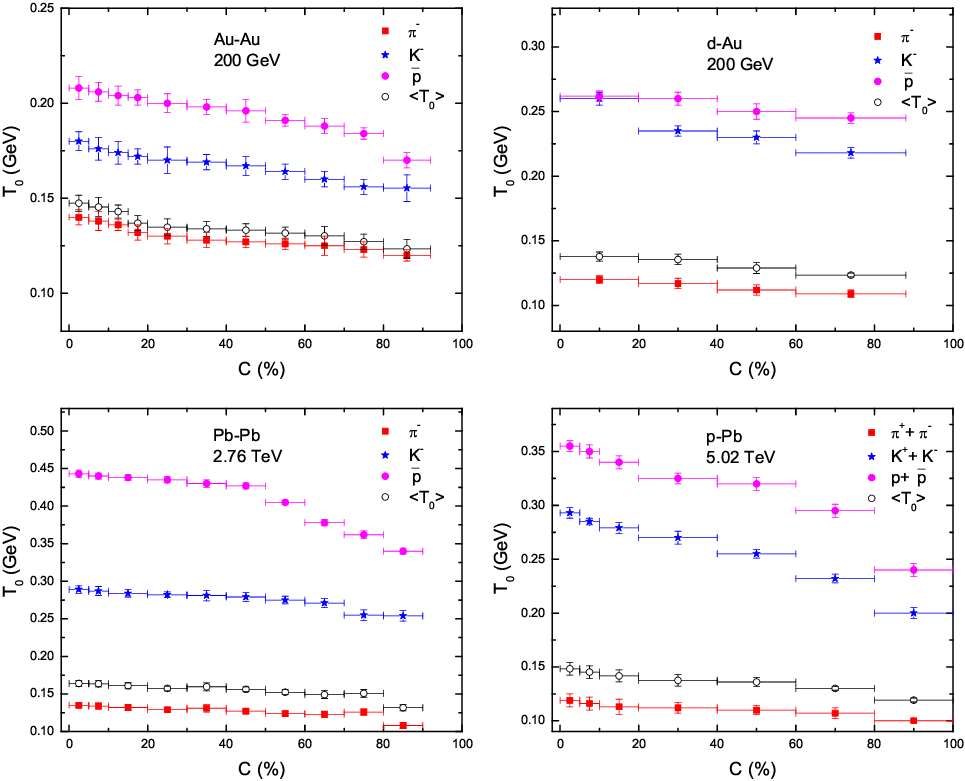}
\end{center}
Fig. 3. Dependence of $T_0$ on event centrality percentage in 200
GeV Au-Au (left-upper), 200 GeV $d$-Au (right-upper), 2.76 TeV
Pb-Pb (left-lower), and 5.02 TeV $p$-Pb (right-lower) collisions.
Different symbols represent different parameter values listed in
Tables 1 and 2. The weighted averages, $\langle T_0\rangle$, over
different particles are shown in the figure together.
\end{figure*}

\begin{figure*}[!htb]
\vskip-0cm
\begin{center}
\includegraphics[width=16.cm]{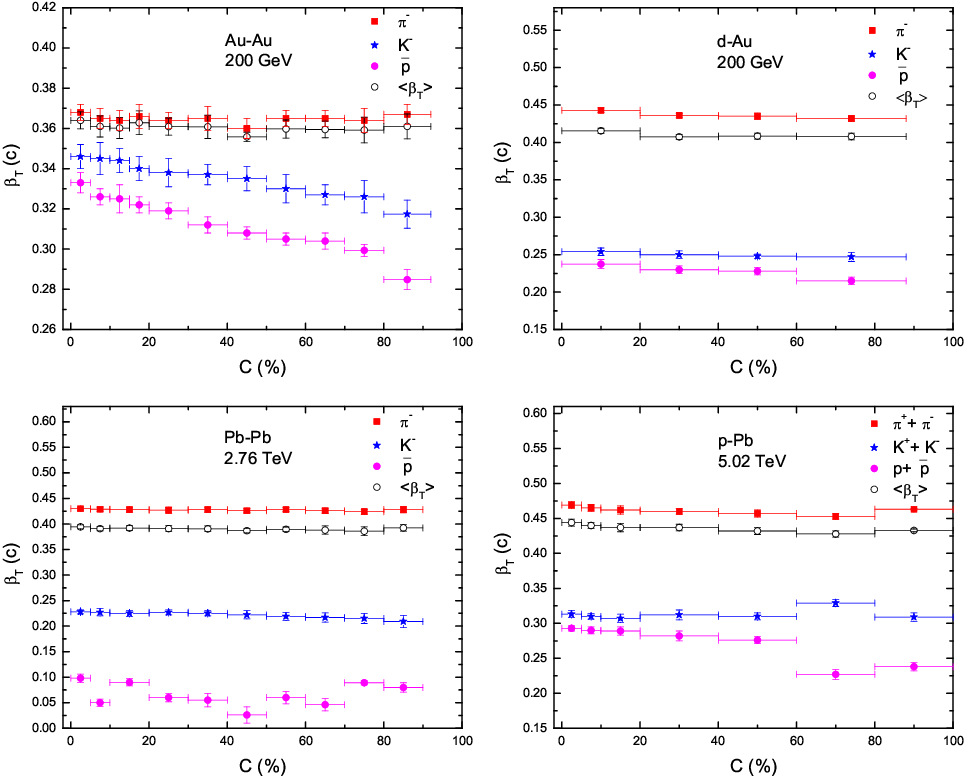}
\end{center}
Fig. 4. Dependence of $\beta_T$ on event centrality percentage in
200 GeV Au-Au (left-upper), 200 GeV $d$-Au (right-upper), 2.76 TeV
Pb-Pb (left-lower), and 5.02 TeV $p$-Pb (right-lower) collisions.
Different symbols represent different parameter values listed in
Tables 1 and 2. The weighted averages, $\langle \beta_T \rangle$,
over different particles are shown in the figure together.
\end{figure*}

In the fit process for Figs. 1 and 2, the parameters $T_0$ and
$\beta_T$ are correlated. In some cases, a larger $T_0$ and a
smaller $\beta_T$ can lead to a similar result by using a smaller
$T_0$ and a larger $\beta_T$ due to the influence of $p_T$ range
and $n_0$ if changeable. To reduce the effect of correlation, we
analyze the mean $p_T$ ($\langle p_T\rangle$) and the
root-mean-square $p_T$ ($\sqrt{\langle p_T^2\rangle}$) over
$\sqrt{2}$ ($\sqrt{\langle p_T^2\rangle/2}$) in Figs. 5 and 6
respectively, which are calculated from the fit function over a
given $p_T$ range of 0--4.5 GeV/$c$, where $T_i$ represents
$\sqrt{\langle p_T^2\rangle/2}$ to denote the initial temperature
of the interacting system according to the color string
percolation model~\cite{37,38,39}. In particular, the weighted
averages, $\overline{\langle p_T\rangle}$ ($\langle T_i\rangle$),
of $\langle p_T\rangle$ ($T_i$) over different particles are shown
in Fig. 5 (6) by the open circles, which are calculated from the
fit function and weighted by yields of different particles. One
can see that $\langle p_T\rangle$ and $T_i$ increase with the
increases of event centrality, collision energy, and particle
mass. With the increase of projectile size in the case of using
the same target nucleus, $\langle p_T\rangle$ and $T_i$ do not
change obviously.

It should be noted that we have used $T_i$ according to
refs.~\cite{37,38,39}. When we use $\langle T_i\rangle$, it is
independent of specie of the measured particle. It is noteworthy
to measure $T_i$ for the emission of different particles in order
to obtain $\langle T_i\rangle$. Although $T_i$ is directly equal
to $\sqrt{\langle p_T^2\rangle/2}$ which can be obtained from
$p_T$ spectra, one should obtain $T_i$ as usual to see its trend.
Our discussions on $T_i$ and $\langle T_i\rangle$ are useful to
understand the excitation degree of the system in the initial
state. Meanwhile, we may compare $T_i$ and $\langle T_i\rangle$
with $T_0$ and $\langle T_0\rangle$ to see the decrease of
temperature in the system evolution.

\begin{figure*}[!htb]
\vskip-0cm
\begin{center}
\includegraphics[width=16.cm]{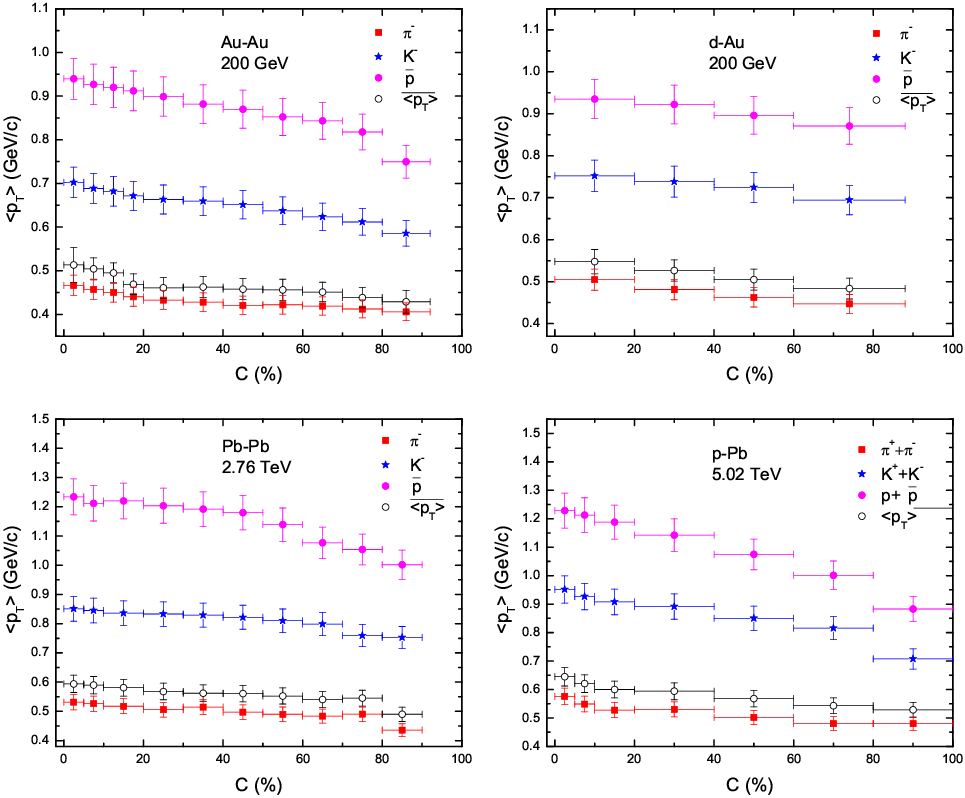}
\end{center}
Fig. 5. Dependence of $\langle p_T\rangle$ on event centrality
percentage in 200 GeV Au-Au (left-upper), 200 GeV $d$-Au
(right-upper), 2.76 TeV Pb-Pb (left-lower), and 5.02 TeV $p$-Pb
(right-lower) collisions. Different symbols represent the results
for different particles based on the parameter values listed in
Tables 1 and 2. The weighted averages, $\overline{\langle
p_T\rangle}$, over different particles are shown in the figure
together.
\end{figure*}

\begin{figure*}[!htb]
\vskip-0cm
\begin{center}
\includegraphics[width=16.cm]{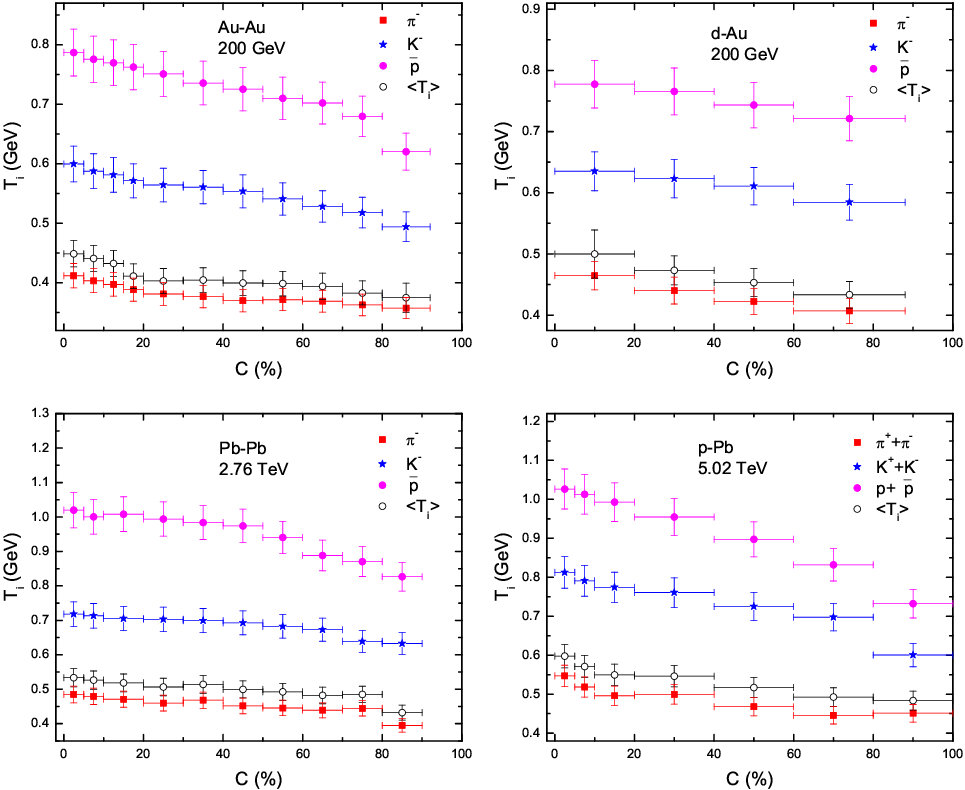}
\end{center}
Fig. 6. Dependence of $T_i$ on event centrality percentage in 200
GeV Au-Au (left-upper), 200 GeV $d$-Au (right-upper), 2.76 TeV
Pb-Pb (left-lower), and 5.02 TeV $p$-Pb (right-lower) collisions.
Different symbols represent the results for different particles
based on the parameter values listed in Tables 1 and 2. The
weighted averages, $\langle T_i \rangle$, over different particles
are shown in the figure together.
\end{figure*}

Generally, large $T_0$ ($\beta_T$) renders wide $p_T$ spectrum and
then large $\langle p_T\rangle$. In central collisions, we have
obtained larger $T_0$ ($\beta_T$) and larger $\langle p_T\rangle$
than those in peripheral collisions. The normalizations ($N_0$)
listed in Tables 1 and 2 are in fact the (pseudo)rapidity density
($dN/d\eta$ or $dN/dy$) of identified particles at mid-rapidity.
One can see a decrease trend from central to peripheral
collisions. This trend is similar to that of $T_0$ ($\beta_T$) and
$\langle p_T\rangle$, but with different slopes. This trend
renders more energy deposition and more violent impact (lager
squeeze) in central collisions, which results in higher excitation
(larger $T_0$) and quicker expansion (larger $\beta_T$).
Naturally, we can obtain larger $\langle p_T\rangle$, $T_i$, and
$dN/d\eta$ in central collisions comparing with those in
peripheral collisions. Although the parameters $T_0$, $\beta_T$,
and $T_i$ are extracted from a model, they are in fact based on
the fit to experimental data. In particular, $T_i$ can be regarded
as an experimental result. It is regretful that there is no direct
experimental values on $T_0$ and $\beta_T$.

It should be noted that $T_i$ used in
refs.~\cite{33,34,35,36,37,38,39} depends on $\sqrt{\langle
p_T^2\rangle/2}$ itself, but not models. If we use the data
directly, we should also obtain $T_i$. It is suitable that $T_i$
is used in the present work. In addition, $T_0$ obtained from
proton spectra in Pb-Pb collisions at the LHC comes out larger
than 300 MeV, in central Pb-Pb collisions this is even 443 MeV.
This way is above the hadronization temperature, which seems
hardly to be understandable. In fact, the weighted average
temperatures $\langle T_0\rangle$ are the kinetic freeze-out
temperatures of the system. The hadronization temperature is
mainly determined by that from pion spectrum due to the fact that
the yield of pions is the largest one among those of the produced
particles. In our opinion, as the kinetic freeze-out temperatures,
the values of $\langle T_0\rangle$ obtained in the present work
are normal. As a statistical quantity and a reflection of mean
thermal motion, $\langle T_0\rangle$ can be naturally used in our
study.

We would like to point out that, although Eq. (5) for the
blast-wave model is obtained by assuming that the system is in
local thermodynamic equilibrium, and therefore, it assumes a
single $T_0$ and $\beta_T$, the single $T_0$ and $\beta_T$ can be
$\langle T_0\rangle$ and $\langle \beta_T\rangle$ respectively.
Generally, $\langle T_0\rangle$ ($\langle \beta_T\rangle$,
$\overline{\langle p_T\rangle}$, or $\langle T_i\rangle$) is
mainly determined by that emitting pions due to the largest yield
of pions at the top RHIC and LHC energies. Meanwhile, $\langle
T_0\rangle$ ($\langle \beta_T\rangle$) reflects approximately a
combined fit to different particle species. Indeed, statistical
fluctuations in the data in different collisions with different
centralities produce fluctuations in the fitted parameters.

The reason that $T_0$ or $\beta_T$ does not decrease with the
increase of event centrality and collision energy renders that the
violent degree of thermal excitation and collective behavior at
the stage of kinetic freeze-out in interacting system does not
decrease with increase of event centrality and collision energy.
This results in faster or the same thermal motion and collective
expansion which are reflected by $T_0$ and $\beta_T$ respectively.
In addition, in peripheral collisions, a larger fraction in high
$p_T$ region is observed due to larger cascade scattering
happening in spectator nucleons and appreciable contribution from
the hard component, and that results in the departure of
statistical law from the thermal model in high $p_T$ region, which
results in the appearance of special $p_T$ range beyond which the
thermal model does not work. In fact, the single thermal model
works generally in low $p_T$ region. For the spectra in high $p_T$
region, we need another thermal component with high temperature,
or the Hagedorn function~\cite{14,15,23,24,25} or its
revisions~\cite{27,28,29,30,31}.

Although the thermal model does not work in the region beyond the
special $p_T$ range, the Hagedorn function~\cite{14,15}, that is
the inverse power-law~\cite{23,24,25} which is based on the
quantum chromodynamics (QCD) theory, can be used to describe the
spectra beyond the special $p_T$ range. Because of the focus of
the present work being the study of event centrality dependence of
$T_0$ and $\beta_T$, we give up to describe the spectra beyond the
special $p_T$ range by using the Hagedorn function~\cite{14,15}.
Based on different pictures in physics, we can use different
methods to describe the same $p_T$ spectra. Different methods are
expected to show similar or reconcilable results if only the
spectra in the special $p_T$ range are
considered~\cite{1,2,3,4,5,6,7,8,9,10,11,12,13}.

In fact, we have another method to describe the spectra beyond the
special $p_T$ range. That is, a two-component thermal model in
which the first component describes the spectra in the special
$p_T$ range and the second one describes the spectra beyond the
special $p_T$ range. As the fraction beyond the special $p_T$
range is small, the two-component thermal model causes a small
increase in $T_0$ and/or $\beta_T$. Because of the increase in
$T_0$ and/or $\beta_T$ being small in the two-component thermal
model, we neglect the contribution of the second component in the
extraction of $T_0$ and $\beta_T$, though the application of the
two-component thermal model is expected to be more successful than
that of the single thermal model.

The observed $\beta_T$ increases slightly or does not change
obviously with the increase of event centrality and collision
energy. This is in agreement with most of the
literature~\cite{1,2,3,4}, though the concrete values are
different from each other. The observed $T_0$ increases slightly
or does not change obviously with the increase of event centrality
and collision energy. This is inconsistent with some
literature~\cite{1,2,3,18,36} and in agreement with
others~\cite{4,16,40,41,42}. In particular, ref.~\cite{43} shows
that $T_0$ extracted from pion spectra in central collisions is
smaller than that in peripheral collisions, and that extracted
from kaon or proton spectra does not depend on the centrality.
Meanwhile, $\beta_T$ in central collisions is larger. Our
observation on $\beta_T$ is in agreement with ref.~\cite{43}. In
fact, due to the anti-correlation between $T_0$ and $\beta_T$ for
a given spectrum, the acceptable sets of free parameter values are
not unique. This renders that the collision process is complex and
more analysis is needed. Indeed, the present work uses
particle-independent $p_T$ range, fixed flow profile ($n_0=2$ as
in ref.~\cite{1}), and small $\beta_T$ change, which results in
different $T_0$ trend with some literature~\cite{2,18}, though the
same blast-wave model is used.
\\

{\subsection{Further discussion}}

The above comparison renders that the event centrality and
collision energy dependent $T_0$ ($\beta_T$) is complex if one
fits the $p_T$ spectra by the same or similar model, let alone
different models. In particular, there exists lots of other models
that go far beyond a blast wave approximation to describe the
$p_T$ spectra~\cite{44,45}, which are not always suitable to
extract $T_0$ and $\beta_T$. In addition, some models do not
extract directly $T_0$ and
$\beta_T$~\cite{2,5,6,7,8,9,10,11,12,13}, though they describe the
$p_T$ spectra to extract the so-called effective temperature, $T$.
Alternatively, $T_0$ is regarded as the intercept in the linear
relation of $T$ against $m_0$, and $\beta_T$ is regarded as the
slope in the linear relation of $\langle p_T\rangle$ against mean
energy (i.e. the mean moving mass $\overline{m}$).

In our opinion, if higher $T_0$ in central collisions and at the
LHC signifies higher excitation degree of interacting system,
lower $T_0$ in central collisions and at the LHC signifies longer
lifetime of hot and dense matter in the case of considering higher
excitation degree. Based on different pictures and functions, the
values and tendency of $T_0$ ($\beta_T$) extracted from the same
$p_T$ spectra are possibly different from each other. The pictures
and functions used the Boltzmann-Gibbs (Fermi-Dirac or
Bose-Einstein) statistics and standard distribution have more
potentials to be the unified `thermometer" and/or ``speedometer"
because they are the most similar to the ideal gas model in
thermodynamics.

It should be noted that the above discussions are in the low $p_T$
region. If we increase the region by a few GeV/$c$, the
conclusions do not change obviously due to the fact that the
values of $T_0$ and $\beta_T$ do not change obviously. In fact,
the fraction of particles with high $p_T$ is very small. Including
particles with high $p_T$ causes a little ($<5\%$) increase in
$T_0$ and/or $\beta_T$, where the $<5\%$ statement is estimated by
us due to little yield of particles from high $p_T$ region. While
decreasing $p_T$ region by a few GeV/$c$ causes obvious change in
$T_0$ and $\beta_T$ due to the large change in fraction. As the
parameters in the blast-wave model, $T_0$ and $\beta_T$ are
sensitive to $p_T$ region contributed by the soft process and
insensitive to $p_T$ region contributed by the hard process.

To extract $T_0$ and $\beta_T$ as accurately as possible, one may
use some restrictions in the fit process. For example, the $p_T$
region does not need to be very wide due to the fact that the
fraction of particles with high $p_T$ is very small. The rapidity
should be in the central region. If the rapidity is in the
fragmentation region, we may transform it to the central region so
that the kinetic energy of directional movement can be eliminated
from the total energy. The particles should be light flavor
particles due to the heavy flavor particles being produced mainly
in the non-thermal process. Anyhow, the same $p_T$ region should
be used in different centralities for a given particle and system
to obtain accurate results as far as possible.

No matter how large the correlation of $T_0$ and $\beta_T$ is,
$\langle p_T\rangle$ and $T_i$ are independent of models and
parameters if the models fit well the data. In fact, $\langle
p_T\rangle$ and $T_i$ are reflections of data sample itself in
case if the models fit the data well. The results will be similar
to Figs. 5 and 6 if we use other models to fit the data well.
Considering two nucleons or partons taking part in each binary
collision, the contribution fraction of each nucleon or parton to
$\langle p_T\rangle$ is 1/2. In the Erlang distribution, $\langle
p_T\rangle/2$ is regarded as the temperature parameter~\cite{46}.
If the contribution fraction of thermal motion to $\langle
p_T\rangle/2$ is $k_0$, we can obtain $T_0$ to be $k_0\langle
p_T\rangle/2$ and $\beta_T$ to be $(1-k_0)\langle
p_T\rangle/2m_0\overline{\gamma}$, where $\overline{\gamma}$
denotes the mean Lorentz factor of the given particles in the
source rest frame. Thus, we can obtain the similar trends of $T_0$
($\beta_T$) to the blast-wave model.

The value of $\langle T_0\rangle$ ($\langle \beta_T\rangle$,
$\overline{\langle p_T\rangle}$, or $\langle T_i\rangle$) for
different particles in Au-Au and $d$-Au collisions, as well as in
Pb-Pb and $p$-Pb collisions, are similar to each other. These
results confirm the maximum size dependent effect~\cite{33}, which
states that the main parameters such as the kinetic freeze-out
temperature and transverse flow velocity are mainly determined by
the heaviest nucleus from proton-nucleus to nucleus-nucleus
collisions. In the maximum size dependent effect, the number of
participant nucleons in collisions of single projectile proton
with target nucleus plays a main role. This renders that central
(peripheral) proton-nucleus and nucleus-nucleus collisions result
in similar results to each other, and peripheral proton-nucleus
and nucleus-nucleus collisions also result in similar results to
proton-proton collisions.

Comparing with our previous work~\cite{33} which studies more
models, the progress in this paper is obvious. In fact, the
dependences of $T_0$, $\beta_T$, $\langle p_T\rangle$, and $T_i$
on centrality are studied in this paper in whole centrality range.
We confirm here that these dependences decrease monotonously from
central to peripheral collisions, which is not always the case for
$T_0$ in some cases if we use particle dependent $p_T$ region and
centrality dependent $n_0$ as used in current
literature~\cite{1,2,3,4,41}. The ``standard pictures" extracted
from low to the highest energies in nuclear collisions show that
$T_0$ increase from central to peripheral collisions. The
difference in $T_0$ between our result and the ``standard picture"
is caused by the choice of $p_T$ range and $n_0$. In our opinion,
$T_0$ and $\beta_T$ depend on particle dependent $p_T$ range and
centrality dependent $n_0$. We have used the fixed $p_T$ range and
invariable $n_0$ in this paper which results in different trend
for $T_0$ from the ``standard picture". It is hard to say that
which picture is more suitable.

Generally, at given collision energy and in given event
centrality, larger system produces more particles. For a given
collision system, with the increase of centrality, larger amount
of particles are produced, and both $T_0$ and $\beta_0$ also
become larger. It is expected that, not only for large system
(e.g. Pb-Pb) but also for small system (e.g. $p$-Pb), the
centrality classes that have similar average $dN/d\eta$ (or $N_0$)
should have similar $T_0$ and $\beta_0$. This paper confirms that
medium and peripheral Pb-Pb collisions that produce similar amount
of particles as central and medium $p$-Pb collisions also have
similar $T_0$ and $\beta_T$. That is to say that, in central and
medium $p$-Pb collisions, both $T_0$ and $\beta_T$ are larger as
those in medium and peripheral Pb-Pb collisions.

Before summary and conclusions, we would like to point out that
the present work is a more systematic study on the dependence of
$T_0$ ($\beta_T$, $\langle p_T\rangle$, or $T_i$) on centrality in
Au-Au, $d$-Au, Pb-Pb, and $p$-Pb collisions at the RHIC and LHC by
using the blast-wave model with Boltzmann-Gibbs statistics, though
our previous work~\cite{47} studied $T_0$ ($\beta_T$, $q$ (entropy
index), $p_0$, and $n$) in central and peripheral Au-Au and Pb-Pb
collisions by using an improved Tsallis distribution~\cite{35,48}.
With regard to the relative size of $T_0$ ($\beta_T$) in central
and peripheral collisions, the results from the two works are
consistent with each other. However, the improved Tsallis
distribution results in larger $\beta_T$ which is inconsistent
with the blast-wave model with Boltzmann-Gibbs statistics or with
Tsallis statistics~\cite{50}. Thus, we did not use the improved
Tsallis distribution in the present work. In addition, two
components were used in our previous work~\cite{47} which resulted
in good fit within and beyond the special $p_T$ range, which is
not the case in the present work which uses one component in fact.

In the above discussion on extracting $T_0$ and $\beta_T$, the
influence of resonances if available are naturally included in the
soft component. As discussed in ref.~\cite{1}, the resonances
change the slopes of the $p_T$ spectra, which affect the values of
$T_0$ and $\beta_T$. In some cases, the spectra in very low $p_T$
region are not available in experiments, which results in larger
$T_0$ and $\beta_T$ in the present work. Although the influence of
resonances on particle yields are not too large~\cite{51}, this
influence on pions is the largest among the considered
particles~\cite{51,52}. To include the influence will result in
lower $T_0$ and $\beta_T$ for pions than for kaons and protons.
This will strengthen further our conclusion on multiple kinetic
freeze-out scenario in detailed analysis. Although the chemical
freeze-out is single scenario in the traditional statistical
thermal model~\cite{52,53,54,55}, the kinetic freeze-out is
possibly mass dependent~\cite{10,22,34,35}. In particular, in a
very recent work~\cite{56}, the two-scenario of chemical
freeze-out is studied and the inclusion of additional resonances
are not sufficient to close the gap between the chemical
freeze-out temperatures for emissions of light and strange
hadrons.

At the end of this discussion, we would like to point out that the
value of $T_0$ obtained from the analysis (for heavier hadrons) is
more than the value of critical temperature ($T_c\approx170$--200
MeV) obtained in lattice QCD calculations~\cite{57,58,59}. This
result does not means that the kinetic freeze-out takes place
before the phase transition. In fact, the value of $\langle
T_0\rangle$ is less than the value of $T_c$, if we consider
averagely and generally. In addition, it is indeed that $T_0$ is a
``true" kinetic freeze-out temperature but not the inverse slope
of the spectra because the transverse flow velocity $\beta_T$ is
extracted out. However, $T_0$ and $T_c$ are obtained from
different ``thermometers". Meanwhile, the larger $T_i$ for heavier
hadrons emission comparing with that from other
methods~\cite{60,61,62,63} is also caused by different
``thermometers", though $\langle T_i\rangle$ is nearly the same as
or close to refs.~\cite{61,62,63}. Before giving a quantitative
comparison for different temperatures, one has to define a unified
``thermometer". This topic is beyond the focus of this paper. We
shall not discuss it further.
\\

{\section{Summary and conclusions}}

We summarize here our main observations and conclusions.

The centrality-dependent double-differential transverse momentum
spectra of charged pions and kaons and (anti)protons produced in
mid-(pseudo)rapidity interval in $\sqrt{s_{NN}}=200$ GeV Au-Au and
$d$-Au, 2.76 TeV Pb-Pb, and 5.02 TeV $p$-Pb collisions are
analyzed by the blast-wave model with Boltzmann-Gibbs statistics.
The model results are approximately in agreement with the
experimental data in special transverse momentum ranges measured
by the PHENIX and ALICE Collaborations.

There are special transverse momentum ranges in some transverse
momentum spectra. The special transverse momentum range increases
from $0\sim2$--3 GeV/$c$ to $0\sim4.5$ GeV/$c$ or a little more
when the event centrality increases from periphery to center. This
range for the strange particle is narrower than that for the
non-strange particle. The dependence of this range on collision
energy is not obvious. The special transverse momentum ranges
appear due to different fractions of participant nucleons in
events with different centralities.

The kinetic freeze-out temperature and the transverse flow
velocity increase slightly in some cases or do not change
obviously in other cases with the increase of event centrality and
collision energy. These outcomes result in faster or the same
thermal motion and collective expansion in central collisions, at
the LHC, and for large system. Comparing with central collisions,
a large fraction in high transverse momentum region is observed in
peripheral collisions due to large cascade scattering happen in
spectator nucleons. The single thermal model does not describe
simultaneously the spectra in both the low and high transverse
momentum regions, though the two-component thermal model is
expected to describe simultaneously the spectra in the two
regions.

The average transverse momentum and initial temperature increase
with the increase of event centrality, collision energy, and
particle mass. With the increase of projectile size in the case of
using the same target nucleus, the two quantities and main
parameters (kinetic freeze-out temperature and transverse flow
velocity) do not change obviously. This confirms the maximum size
dependent effect, which states that the main parameters such as
the kinetic freeze-out temperature and transverse flow velocity
are mainly determined by the heaviest nucleus from proton-nucleus
to nucleus-nucleus collisions.

The present work also confirms the multiple kinetic freeze-out
scenario if we use the detailed analysis. Although the chemical
freeze-out is single scenario in the traditional statistical
thermal model, there is also two-scenario of chemical freeze-out
studied in literature. The present work shows that the kinetic
freeze-out is possibly particle mass dependent, which show an
increase of kinetic freeze-out temperature with the increase of
particle mass. If the influence of resonances can be measured in
experiments in detail, the multiple kinetic freeze-out scenario is
expected to strengthen further.
\\
\\
{\bf Acknowledgments}

This work was supported by the National Natural Science Foundation
of China under Grant Nos. 11575103 and 11947418, the Chinese
Government Scholarship (China Scholarship Council), the Scientific
and Technological Innovation Programs of Higher Education
Institutions in Shanxi (STIP) under Grant No. 201802017, the
Shanxi Provincial Natural Science Foundation under Grant No.
201901D111043, and the Fund for Shanxi ``1331 Project" Key
Subjects Construction.
\\
\\
{\bf Data availability}

The data used to support the findings of this study are included
within the article and are cited at relevant places within the
text as references.
\\
\\
{\bf Compliance with ethical standards}
\\
\\
{\bf Conflict of interest}

The authors declare that there are no conflicts of interest
regarding the publication of this paper. The funding agencies have
no role in the design of the study; in the collection, analysis,
or interpretation of the data; in the writing of the manuscript,
or in the decision to publish the results.
\\
\\

\end{multicols}

{\small
\begin{thebibliography}{99}
\setlength{\itemsep}{-1pt}

\bibitem{1}
E Schnedermann, J Sollfrank and U Heinz {\it Phys. Rev. C} {\bf
48} 2462 (1993)

\bibitem{2}
B I Abelev et al. [STAR Collaboration] {\it Phys. Rev. C} {\bf 79}
034909 (2009)

\bibitem{3}
B I Abelev et al. [STAR Collaboration] {\it Phys. Rev. C} {\bf 81}
024911 (2010)

\bibitem{4}
Z B Tang, Y C Xu, L J Ruan, G van Buren, F Q Wang and Z B Xu {\it
Phys. Rev. C} {\bf 79} 051901(R) (2009)

\bibitem{5}
S Takeuchi, K Murase, T Hirano, P Huovinen and Y Nara {\it Phys.
Rev. C} {\bf 92} 044907 (2015)

\bibitem{6}
H Heiselberg, A M Levy {\it Phys. Rev. C} {\bf 59} 2716 (1999)

\bibitem{7}
U W Heinz, arXiv:hep-ph/0407360 (2004)

\bibitem{8}
R Russo {\it PhD Thesis} (Universita degli Studi di Torino, Italy)
(2015), arXiv:1511.04380 [nucl-ex] (2015)

\bibitem{9}
H-R Wei, F-H Liu and R A Lacey {\it Eur. Phys. J. A} {\bf 52} 102
(2016)

\bibitem{10}
H-L Lao, H-R Wei, F-H Liu and R A Lacey {\it Eur. Phys. J. A} {\bf
52} 203 (2016)

\bibitem{11}
H-R Wei, F-H Liu and R A Lacey {\it J. Phys. G} {\bf 43} 125102
(2016)

\bibitem{12}
J Cleymans and D Worku {\it Eur. Phys. J. A} {\bf 48} 160 (2012)

\bibitem{13}
H Zheng and L L Zhu {\it Adv. High Energy Phys.} {\bf 2016}
9632126 (2016)

\bibitem{14}
R Hagedorn {\it Riv. Nuovo Cimento} {\bf 6}(10), 1 (1983)

\bibitem{15}
B Abelev et al. [ALICE Collaboration] {\it Eur. Phys. J. C} {\bf
75} 1 (2015)

\bibitem{16}
S S Adler et al. [PHENIX Collaboration] {\it Phys. Rev. C} {\bf
69} 034909 (2004)

\bibitem{17}
A Adare et al. [PHENIX Collaboration] {\it Phys. Rev. C} {\bf 88}
024906 (2013)

\bibitem{18}
B Abelev et al. [ALICE Collaboration] {\it Phys. Rev. C} {\bf 88}
044910 (2013)

\bibitem{19}
B Abelev et al. [ALICE Collaboration] {\it Phys. Lett. B} {\bf
728} 25 (2014)

\bibitem{20}
S Chatrchyan et al. [CMS Collaboration] {\it Eur. Phys. J. C} {\bf
72} 1945 (2012)

\bibitem{21}
M K Suleymanov {\it Int. J. Mod. Phys. E} {\bf 27} 1850008 (2018)

\bibitem{22}
M Waqas, F-H Liu, S Fakhraddin, M A Rahim {\it Indian J. Phys.}
{\bf 93} 1329 (2019)

\bibitem{23}
R Odorico {\it Phys. Lett. B} {\bf 118} 151 (1982)

\bibitem{24}
G Arnison et al. [UA1 Collaboration] {\it Phys. Lett. B} {\bf 118}
167 (1982)

\bibitem{25}
T Mizoguchi, M Biyajima and N Suzuki {\it Int. J. Mod. Phys. A}
{\bf 32} 1750057 (2017)

\bibitem{26}
K Aamodt et al. [ALICE Collaboration] {\it Phys. Lett. B} {\bf
693} 53 (2010)

\bibitem{27}
A De Falco [for the ALICE Collaboration]. {\it J. Phys. G} {\bf
38} 124083 (2011)

\bibitem{28}
I Abt et al. [HERA-B Collaboration] {\it Eur. Phys. J. C} {\bf 50}
315 (2007)

\bibitem{29}
B Abelev et al. [ALICE Collaboration] {\it Phys. Lett. B} {\bf
710} 557 (2012)

\bibitem{30}
B Abelev et al. [ALICE Collaboration] {\it Phys. Lett. B} {\bf
718} 295 (2012) and Corrigendum {\it Phys. Lett. B} {\bf 748} 472
(2015)

\bibitem{31}
I Lakomov [for the ALICE Collaboration] {\it Nucl. Phys. A} {\bf
931} 1179 (2014)

\bibitem{32}
B Abelev et al. [ALICE Collaboration] {\it Phys. Lett. B} {\bf
708} 265 (2012)

\bibitem{33}
H-L Lao, F-H Liu, B-C Li, M-Y Duan and R A Lacey {\it Nucl. Sci.
Tech.} {\bf 29} 164 (2018)

\bibitem{33a}
P Bozek, {\it AIP Conf. Proc.} {\bf 1422} 34 (2012)

\bibitem{33b}
M Alqahtani, D Almaalol, M Nopoush, R Ryblewski, M Strickland,
{\it Nucl. Phys. A} {\bf 982} 423 (2019)

\bibitem{33c}
Y-L Yan, Y Cheng, D-M Zhou, B-G Dong, X Cai, B-H Sa, L P Csernai,
{\it J. Phys. G} {\bf 40} 025102 (2013)

\bibitem{34}
S Chatterjee and B Mohanty {\it Phys. Rev. C} {\bf 90} 034908
(2014)

\bibitem{35}
D Thakur, S Tripathy, P Garg, R Sahoo and J Cleymans {\it Adv.
High Energy Phys.} {\bf 2016} 4149352 (2016)

\bibitem{35a}
R Sahoo {\it Association of Asia Pacific Physical Societies
Bulletin} {\bf 29}(4), 16 (2019)

\bibitem{36}
I Melo and B Tom{\'a}{\v s}ik {\it J. Phys. G} {\bf 43} 015102
(2016)

\bibitem{37}
L J Gutay, A S Hirsch, R P Scharenberg, B K Srivastava and C
Pajares {\it Int. J. Mod. Phys. E} {\bf 24} 1550101 (2015)

\bibitem{38}
A S Hirsch, C Pajares, R P Scharenberg and B K Srivastava {\it
Phys. Rev. D} {\bf 100} 114040 (2019)

\bibitem{39}
P Sahoo, S De, S K Tiwari and R Sahoo {\it Eur. Phys. J. A} {\bf
54} 136 (2018)

\bibitem{40}
S Chatterjee, B Mohanty and R Singh {\it Phys. Rev. C} {\bf 92}
024917 (2015)

\bibitem{41}
Z B Tang, L Yi, L J Ruan, M Shao, C Li, H F Chen, B Mohanty and Z
B Xu {\it Chin. Phys. Lett.} {\bf 30} 031201 (2013)

\bibitem{42}
D Thakur, S Tripathy, P Garg, R Sahoo and J Cleymans, Proceedings
of the 11th Workshop on Particle Correlations and Femtoscopy
(WPCF2015), 3--7 Nov. 2015, Warsaw, Poland, arXiv:1603.04971
[hep-ph] (2016)

\bibitem{43}
B De {\it Eur. Phys. J. A} {\bf 50} 138 (2014)

\bibitem{44}
N Armesto, N Borghini, S Jeon et al. (editors), S Abreu, S V
Akkelin, J Alam et al. (authors) {\it J. Phys. G} {\bf 35} 054001
(2008)

\bibitem{45}
S A Bass, M Bleicher, W Cassing, A Dumitru, H J Drescher, K
Eskola, M Gyulassy, D Kharzeev, Y Kovchegov, Z Lin, D Molnar, J Y
Ollitrault, S Pratt, J Rafelski, R Rapp, D Rischke, J
Schaffner-Bielich, B Schlei, A Snigerev, H Sorge, D Srivastava, J
Stachel, D Teaney, R Thews, S Vance, I Vitev, R Vogt, X N Wang, B
Zhang and J Zim{\'a}nyi {\it Nucl. Phys. A} {\bf 661} 205 (1999)

\bibitem{46}
W-J Xie {\it Chin. Phys. C} {\bf 35} 1111 (2011)

\bibitem{47}
H-L Lao, F-H Liu and R A Lacey {\it Eur. Phys. J. A} {\bf 53} 44
(2017)

\bibitem{48}
T Bhattacharyya, J Cleymans, A Khuntia, P Pareek and R Sahoo {\it
Eur. Phys. J. A} {\bf 52} 30 (2016)


\bibitem{50}
H-L Lao, F-H Liu, B-C Li and M Y Duan {\it Nucl. Sci. Tech.} {\bf
29} 82 (2018)

\bibitem{51}
N Yu and X F Luo {\it Eur. Phys. J. A} {\bf 55} 26 (2019)

\bibitem{52}
J Cleymans, B K{\"a}mpfer and S Wheaton {\it Phys. Rev. C} {\bf
65} 027901 (2002)

\bibitem{53}
F Becattini, J Manninen and M Ga{\'z}dzicki {\it Phys. Rev. C}
{\bf 73} 044905 (2006)

\bibitem{54}
A Andronic, P Braun-Munzinger, K Redlich and J Stachel {\it Nucl.
Phys. A} {\bf 789} 334 (2007)

\bibitem{55}
J Cleymans, H Oeschler, K Redlich and S Wheaton {\it Phys. Rev. C}
{\bf 73} 034905 (2006)

\bibitem{56}
P Alba, V M Sarti, J Noronha-Hostler, P Parotto, I
Portillo-Vazquez, C Ratti and J M Stafford, {\it Phys. Rev. C}
{\bf 101} 054905 (2020)

\bibitem{57}
M Cheng, N H Christ, S Datta, J van der Heide, C Jung, F Karsch, O
Kaczmarek, E Laermann, R D Mawhinney, C Miao, P Petreczky, K
Petrov, C Schmidt, W Soeldner and T Umeda, {\it Phys. Rev. D} {\bf
77} 014511 (2008)

\bibitem{58}
A Bazavov, T Bhattacharya, M Cheng, N H Christ, C DeTar, S Ejiri,
S Gottlieb, R Gupta, U M Heller, K Huebner, C Jung, F Karsch, E
Laermann, L Levkova, C Miao, R D Mawhinney, P Petreczky, C
Schmidt, R A Soltz, W Soeldner, R Sugar, D Toussaint and P Vranas,
{\it Phys. Rev. D} {\bf 80} 014504 (2009)

\bibitem{59}
Y Aoki, Z Fodor, S D Katz and K K Szabo, {\it J. High Energy
Phys.} {\bf 0601} 089 (2006)

\bibitem{60}
R A Soltz, I Garishvili, M Cheng, B Abelev, A Glenn, J Newby, L A
L Levy and S Pratt, {\it Phys. Rev. C} {\bf 87} 044901 (2013)

\bibitem{61}
M Csan{\'a}d, Proceedings of the 7th Workshop on Particle
Correlations and Femtoscopy (WPCF2011), 20--24 Sept. 2011, Tokyo,
Japan, {\it PoS} {\bf WPCF2011} 035 (2011), arXiv:1202.5974
[nucl-th] (2012)

\bibitem{62}
J K Nayak, J Alam, S Sarkar and B Sinha, {\it J. Phys. G} {\bf 35}
104161 (2008)

\bibitem{63}
D K Srivastava, R Chatterjee and M G Mustafa, arXiv:1609.06496
[nucl-th] (2016)

\end{thebibliography}}
\end{document}